\documentclass[12pt]{article}
\usepackage{amssymb}
\usepackage{amsmath}

\def\beq{\begin{equation}}\def\eeq{\end{equation}}
\def\bea{\begin{eqnarray}}\def\eea{\end{eqnarray}}

\begin{document}

\title{Completely local interpretation of quantum field theory}
 
\author{Roman Sverdlov
\\Raman Research Institute,
\\C.V. Raman Avenue, Sadashivanagar, Bangalore --560080, India}
\date{September 5, 2010}
\maketitle
 
\begin{abstract}

The purpose of this paper is to come up with a framework that "converts" existing concepts from configuration space to ordinary one. This is done by modeling our universe as a big "computer" that simulates configuration space. If that "computer" exists in ordinary space and is ran by "classical" laws, our theory becomes "classical" by default. We have first applied this concept to a version of quantum field theory in which elementary particles have size (that is, a theory that does not yet exists). After that, we have also done the same with Pilot Wave model of discrete jumps, due to D\"urr et el. 

\end{abstract}

\subsection*{1. Introduction}

When we think of non-relativistic quantum mechanics of a single particle, we can re-interpret it as a classical wave mechanics: a field $\psi$ "evolves" according to "classical" Schr\"odinger's equation. We do not have to think of it as probability. It simply \emph{happens} that the probability of the "collapse" of that "classical" field is proportional to its amplitude square. Furthermore, Bohm's Pilot Wave model can provide a mechanism by which the probability of finding a particle at any given place is proportional to $\vert \psi \vert^2$ (even though in case of single particle there is no "collapse" since there is no entropy required for decoherence).

However, when we introduce more than one particle, this qualitative picture completely changes. The function $\psi$ is no longer a function on a position space but rather on a \emph{configuration space}, as it assigns complex amplitudes to different configurations of particles; and, in quantum field theory it becomes a function on Fock space, where the numbers of particles are no longer fixed. Since we are used to think of a "field" as a function strictly on $\mathbb{R}^3$, we can no longer view probability amplitudes as one. This forces us to take the word of "probability" in the term "probability amplitude" more seriously, which brings us back to the paradox of its complex value. 

The ultimate answer to this question is to restore the usual three dimensional space. It is possible to do so by the following argument. Suppose we were living in a classical, three dimensional world. In that world, we could have designed a computer program that simulates configuration space, probability amplitudes, and everything else we have to deal with in quantum mechanics. Now, there is no such thing as "computer program". In reality, computer is made out of particles, and what it shows in the screen is a result of complex interaction of particles. This implies that there are \emph{three dimensional} processes taking place that "simulate" non-existant, multidimensional configuration space. In this work we will come up with  one such three dimensional classical device that does it. 

We then take advantage of this philosophy and notice that we can "write a computer program" not only for computing amplitudes of quantum states, but we can also write a computer program for Pilot Wave model as well. For the purposes of this paper, we will "write a program" for the model with Stochastic jumps proposed by D\"urr et el (see \cite{jumps}). We chose this specific approach because it invokes discrete quantum states, and, as we shall see, discrete states are easiest to model that way. But, at the same time, our "computer" is ran by \emph{classical} laws that are \emph{smooth} and deterministic. This cures the model from the violation of determinism due to stochastic nature of originally proposed jumps. In principle, the same can be done for any other Pilot Wave model. But, since the latter is a lot more complicated, we leave other Pilot Wave models for future research. 

It has to be pointed out, however, that while our "computer" is ran by "local" signals, their speed of propagation is much faster than the speed of light. After all, it is not possible to build a computer that simulates signals \emph{faster} than the ones in the world in which it lives. After all, any signal that computer simulates is a result of a set of signals between its particles. However, since our intuition does not demand relativistic covariance, I don't regard it as a big problem. The only things our intuition \emph{does} demand are (non-relativistic) locality and determinism, and the mechanism by which our "computer" operates \emph{is} both  local and deterministic. We then argue that the appearance of relativity in the lab is only a result of the specifics of our Hamiltonian. 
 
On the first glance it might sound like cheating. After all, we could use this argument for anything and everything. Suppose, for example, we didn't like Coulumb's law, and we liked the $1/{r^{10}}$ a lot better. We could then model computer, based on $1/{r^{10}}$ that simulates $1/r^2$ on its screen. To our defense, we will point out that we \emph{always} arrive at the same problem when we are trying to do something we don't know. Suppose, for example, Newton didn't have the numerical information that he used in analysis of Kepler's laws. Then, if he would come up with the same law of gravity, it would look very arbitrary.

From the latter point of view, the ultimate reason why our theory looks arbitrary is simply that we don't have any information of what happens on the small scale, which is what forces us to be creative. While this is not a pleasant thing to hear, it is a lot better than saying that our very classical intuition on that scale is wrong. Of course, that might be the case, but it doesn't \emph{have} to be. The purpose of this work is to show one way in which the processes on small scale \emph{might} look classical. This, of course, is just one way out of many other alternatives, and I do not claim this to be the truth. 

\subsection*{2. How wave function is "encoded"}

According to our model, the configuration space (or Fock space in case of quantum field theory) is discretized. Every discrete state is represented by a subset of $\mathbb{R}^3$. More specifically, we have a fixed configuration of particles in $\mathbb{R}^3$. By specifying a specific subset of $\mathbb{R}^3$ we are choosing to "look at" the particles that are inside that subset, while "ignoring" all other particles. This specifies the quantum state. Now, if we are going to alter the subsets of $\mathbb{R}^3$ we are "looking" at, this will result in a \emph{perception} that the particles are either created or destroyed. Now, if we make sure that every such subset of $\mathbb{R}^3$ is "dense enough" (that is, if small enough neighborhood of every point intersects all of the subsets in our list), then this will allow for a perception of continuous processes, such as motion of particles, on small enough scales. 

 To see how this model works, consider a simple example. Suppose the configuration space has only $3$ elements. We assume that we have one \emph{single} configuration of particles, none of which can move (and, therefore, their common reference frame is a "preferred one"). So, for example, we have $5$ electrons, $4$ protons, and $8$ photons. We assume a toy model of only having one space and one time dimension. So, the $x$ coordinates of $5$ electrons are $4.89$, $1.74$. $6.95$, $5.26$ and $8.31$. The $x$ coordinates of $4$ protons are $3.76$, $2.38$, $6.11$ and $4.75$. Finally, the $x$ coordinates of the photons are $2.98$, $4.64$, $7.22$, $7.23$, $8.11$, $4.87$, $1.39$ and $3.68$.

Now we will break set $\mathbb{R}$ into three subsets, $S_1$, $S_2$ and $S_3$. The set $S_1$ consists of all numbers whose second digid after the "dot" is neither divisible by $2$ nor by $3$. The set $S_2$ consists of all numbers whose second digit after the "dot" is divisible by $2$ but \emph{not} by $3$. Finally, the set $S_3$ consists of all numbers whose second digit after the "dot" is divisible by $3$. In general, the number of such sets is the same as the number of points in configuration space (which, of course, is a very large number). As stated previously, in order to allow the combination of creation and annihilation to create an illusion of continuous motion, these sets have to be defined in such a way that they intersect small enough neighborhoods of each point, and the measures of these intersections are roughly the same. Apart from that, the details of their definitions are not very important. 

Now, to $S_1$, $S_2$ and $S_3$, we associate quantum states $\vert s_1 >$, $\vert s_2 >$ and $\vert s_3 >$, respectively. These states consist of configurations of particles whose coordinates happen to be elements of $S_1$, $S_2$ and $S_3$, respectively. Thus, $\vert s_1>$ consists of two electrons, at locations $6.95$ and $8.31$, two protons, at locations $6.11$ and $4.75$, and two photons, at locations $8.11$ and $4.87$. The $\vert s_2 >$ consists of one electron at a location $1.74$ ($5.26$ is disqualified since $6$ is divisible by $3$), one proton at a location $2.38$, and four photons at locations $2.98$, $4.64$, $7.22$ and $3.68$. Finally, $\vert s_3 >$ consists of $2$ electrons at locations $4.89$ and $5.26$, $1$ proton at a location $3.76$, and two photons at locations $7.23$ and $1.39$. 

Now, we introduce a complex valued field $\psi$ on our space $\mathbb{R}$. We come up with a \emph{local} dynamics of $\psi$ in such a way that \emph{at the equilibrium} $\psi (x) \approx \psi (x')$ as long as both $x$ and $x'$ are elements the same set $S_k$, for some $k$. If such is the case, then we can simply \emph{define} the amplitude of $\vert s_k >$ as a value of $\psi (x)$ for any $x \in S_k$. At the same time, in order for the quantum amplitudes to evolve, that probability amplitude changes with time. This, of course, violates the relativistic covariance, and it logically corresponds to the violation of relativistic covariance in the "standard" definition of configuration space. However, this new picture allows us to continue to view quantum mechanics as a classical field theory in ordinary space, albeit non-relativistic one.  

It should be pointed out that an element of the set $S_k$ that "represents" $\vert s_k >$ does not have to coincide with the locations of any of the particles. Thus, $\psi(57.11)$ is approximately equal to the amplitude of $\vert s_1 >$, even though the point $x=57.11$ is more than $47$ units away from any of the particles in configuration $\vert s_1>$. On the other hand, $\psi (6.94)$ approximates the amplitude of $\vert s_2>$ and \emph{not} $\vert s_1 >$, despite the fact that $x =6.94$ is only $0.01$ units away from the electron at $x=6.95$ which is part of $\vert s_1 >$ state. In other words, \emph{despite} the fact that the dynamics of $\psi$ is local, at the equilibrium $\psi$ might change very rapidly in space, and \emph{yet} have similar values at the points that fall into the same state, no matter how far away they are.

The particles, themselves, do not move. But whenever a "measurement" is performed, $\psi$ collapses into an "extreme state" where it is $1$ at all $x \in S_k$ and $0$ everywhere else. Thus, all the particles whose coordinates are \emph{not} elements of $S_k$ are "hidden from view". Thus, if multiple measurements are performed, different particles get "hidden" each time. If we have two electrons very close to each other, and first electron was hidden at $t_1$ but \emph{not} at $t_2$ while the second one was \emph{not} hidden at $t_1$ but became hidden at $t_2$, that might lead to an \emph{appearance} that electron have moved from one location to another, provided that $t_2-t_1$ is small. In reality, however, both of these electrons were always stationary in their respective locations, and always existed as separate particles.

Let us illustrate a point with our example, of a configuration space consisting of three points. Suppose we performed two separate measurements, at $t=t_1$ and $t=t_2$. We have found that, at $t=t_1$, $\psi (37.91) = \psi (69.32) =0$ and $\psi (107.13) =1$; on the other hand, at $t=t_2$ we got $\psi (25.61) = \psi (273.83) =0$ and $\psi (415.72) =1$. From this we know that at time $t=t_1$ our system had collapsed to $\vert s_3 >$ and at the time $t=t_2$ it had collapsed to $\vert s_2 >$. Thus, at $t_1$ we had one electron at a location $1.74$, one proton at a location $2.38$, and four photons at locations $2.98$, $4.64$, $7.22$ and $3.68$. Then, at $t_2$ we had  $2$ electrons at locations $4.89$ and $5.26$, $1$ proton at a location $3.76$, and two photons at locations $7.23$ and $1.39$. As far as most particles are concerned, we wrongly interpret it as them being created and annihilated. Furthermore, we also wrongly decided that one of the photons have "moved" from location $7.22$ to $7.23$. In reality, none of the particles were created or destroyed, and none were "moving" either. Both illusions were accomplished through particles hiding from our view. 

It should also be pointed out that, no matter how we define different $S_k$, and no matter what the stationary particle configuration is, we have a "global" law that if at least one of the particles "moves", then all have to "move". That is due to the fact in order for at least one particle to move, we have to switch states. But, since every point in space is "reserved" for only one state, once the states were switched, none of the particles can be found at any point "occupied" by the old state. In order for this \emph{not} to contradict our experience,  we have to make sure that any $S_k$ intersects with small enough neighborhood of any given point. In this case, up to some coarse-graining, every point will be \emph{approximately} represented in any given state. Furthermore, we would like to make sure that the intersections with neighborhoods we have just mentioned have roughly the same measure. This allows us to avoid unwanted variations of "densities" of the particles on larger scales. 

It should also be noticed that, due to the fact that particles do not move, in our example we can \emph{not} obtain any states besides the three that we listed. The reason in reality we have every possible configuration of particles is simply because we have a lot more particles, and a lot more $S$-s. Once these numbers are large enough, we can statistically expect to generate an approximation to any conceivable state, in at least one of these $S$-s. Now, statistically we can also expect to generate several $S$-s that simulate the same state. But, again from statistical point of view, we expect this overcounting to be approximately the same for different $S$-s, which allows us to expect the probability amplitudes to be consistent with quantum field theory. 

The main challenge of this work is to obtain the \emph{global} picture we have just outlined thorough the \emph{local} dynamics of $\psi$ and some other fields.  In particular, we would like our \emph{local} theory to predict that, at any given moment in time, $\psi$ can vary a lot within any small neighborhood, and yet it is constant within each set, despite the fact that our sets are "spread out" throughout the whole universe.  We accomplish this goal by introducing two stationary fields, $u_1 (\vec{x})$ and $u_2(\vec{x})$ and comming up with a dynamics that assures us that each point in space receives signals with frequency $u_1 (x)$ and emits the ones with frequency $u_2 (x)$. In this case, the points with similar values of $u_1$ are "listening" to the same wavelength.  

To illustrate the way it works, suppose there are three people, $A$, $B$, and $C$. People $A$ and $B$ are neighbors, while a person $C$ lives at the opposite part of the world. A person $A$ and a person $C$ are both listening to the radio station $1$, while a person $B$ is listening to radio station $2$. If these people are completely isolated from any sources of information other than their respective radio stations, then the views of a person $A$ will be the same as the views of the person $C$ and \emph{not} person $B$, despite the fact that person $B$ is very close to a person $A$, while person $C$ is far away. At the same time, the picture is completely local, since the mechanism by which radio waves spread is a local one. 

Of course, in case of the radio we have amplifiers that assure us that signals don't get weakened in space. Since in our case we don't have these, we instead assume that the universe is compact (for example, a large sphere). In light of the fact that we admit that we violate relativity, we also assume that the speed of radio waves is much faster than the speed of light; in fact, it is so large that they can circle the universe within a very short time. In light of this, our waves circling the universe several times and, as a result, it is no longer important where they were emitted; their intensity, on average, is the same everywhere. Due to the compactness of the universe, their intensity does \emph{not} go to zero. 

Finally it is improtant to address the unnatural behavior of $u_1$ and $u_2$. First of all, as mentioned earlier, they have to be very fast varying. Secondly, in order for the technicalities of the theory to work, they have to be nearly integer most of the time. Both of these assumptions are very unnatural. In order to make the theory more natural, we postulate $v_1$ and $v_2$ as basic fields, both of which are slowly-varying, differentiable functions (and, of course, non-integer most of the time). We then identify $u_1$ and $u_2$ with $g(v_1)$ and $g(v_2)$, respectively, where $g(x)$ has the properties that we would like $u_1$ and $u_2$ to posess. For example, we can set 
\beq g(x) =   \frac{2}{\pi} \sum_{k=-n}^n tan^{-1} (l(m \; sin \; x - \lfloor m \; \sin \; x \rfloor)) \eeq
where $l$, $m$ and $n$ are very large integers. The fact that $l$ is large assures us that each term in the sum is either close to $\pi/2$ or $-\pi/2$. Thus, the factor $2/ \pi$ assures us that each term is close to either $-1$ or $1$. Thus, the overall sum is close to \emph{any} integer between $-2n-1$ and $2n+1$. Thus, in order to get the desired number of points in configuration space, $n$ has to be equal to that number \emph{divided by 2}. Finally, the fact that we have used $m \; sin \; x - \lfloor m \; \sin \; x \rfloor$ instead of $x$ itself assures us that even when $v_1$ and $v_2$ are smoothly increasing (or decreasing), the functions $v_1$ and $v_2$ \emph{still} quickly fluctuate between \emph{all} integer values between $-2n$ and $2n$, as desired. The fact that $m$ is large implies that these fluctuations are quick.  As a result, by setting $u_1 = g(v_1)$ and $u_2 = g(v_2)$ we obtain desired behavior for $u_1$ and $u_2$ while postulating $v_1$ and $v_2$ to be of a lot more natural form. 

\subsection*{3. The Hamiltonian}

In the previous section we have explained the way we define the complex amplitudes in a configuration space as a function in an ordinary, three dimensional one. We have also defined the concept of evolution of our states, and have shown how apparent creation and annihilation of the particles is possible despite the fact that in reality their number is fixed. We have likewise shown how appearance of their motion is possible while they are stationary. All of these are kinematical concepts. In this section we will move to dynamics. In other words, we will explain what we mean by the Hamiltonian that guides the evolution of a wave function.

Again, we are working with a toy model where configuration space has only three points. But this time, in addition to sets $S_1$, $S_2$ and $S_3$, we will introduce the sets $T_1$, $T_2$ and $T_3$. The sets $S_1$, $S_2$ and $S_3$ are defined as before; on the other hand, $T_1$, $T_2$ and $T_3$ are defined based on the \emph{first} digit after the dot rather than the second one; the rest of the definition is the same. So, for example, $3.14$ is an element of $S_2 \cap T_1$. For any $x \in S_k \cap T_l$, we would like $e^{iH(x)}$ to represent the probability amplitude of the transition \emph{from} $\vert s_k >$ \emph{to} $\vert s_l >$. Thus, $e^{iH(3.14)}$ is a probability amplitude of transition from $\vert s_2 >$ to $\vert s_1 >$, while $e^{iH(3.41)}$ is a probability amplitude of transition in the opposite direction. The unitarity condition implies that $H(3.41) = - H (3.14)$. 

Now suppose we only have three electrons, and they are located at $5.31$, $5.32$, and $7.83$. Thus, the state $\vert s_1 >$ corresponds to one electron at $5.31$, the state $\vert s_2 >$ corresponds to one electron at $5.32$ and the state $\vert s_3 >$ corresponds to one electron at $7.83$. From completely local point of view, none of the states can transition to each other, since none of the particles are "touching". Given that, from our setup, it is \emph{always} bound to be the case, we instead say that the transition \emph{can} occur if two particles are close enough (we will discuss this issue in more detail shortly). Thus,  we would like the probability amplitudes of transitions between $\vert s_1 >$ and $\vert s_2 >$ to be large, while the probability amplitudes between $\vert s_3 >$ and any of the other two states to be small. This means that we would like $e^{iH(7.54)}$ and $e^{iH(3.81)}$ to be large, while  having $e^{iH(5.38)}$, $e^{iH(2.63)}$, $e^{iH(5.13)}$ and $e^{iH(9.35)}$ all small. 

Now, suppose we add another electron at $8.72$. Then the probability amplitude of the transition between $S_1$ and $S_2$ will become small, since there is no electron near $8.72$ in $S_1$. This phenomenon can be explained by expressing $H$ as an integral of a \emph{local} function $h(\vec{x})$, 
\beq H = \int h (\vec{x}) d^3 x \eeq
In the above expression, $e^{ih(\vec{x})}$  represents a probabilityamplitude of an imaginary \emph{quasi-local} transition, with some specified "small" scale (in our case, lets say that scale is $0.2$) and $e^{iH}$ on the other hand represents the probability amplitude of the transition between two global states. Since exponent of the sum is a product of exponents, simple algebra tells us that $e^{iH}$ is a product of $e^{ih}$, as desired. 

In our case, despite the fact that the probability amplitude of transition between $S_2$ and $S_3$ are small, $e^{ih (2.43)}$ is large. After all, as far as neighborhood of $2.43$ is concerned, there are no particles in either state and the probability amplitude of transition between empty states is large. At the same time, however, we assign to $H$ pointwise values too, not just to $h$ (namely, for any $x \in S_k$, $H(x) \approx H(S)$). In light of this, $e^{iH (2.43)}$ is small, despite $e^{ih(2.43)}$ being large. This is due to the fact that $e^{ih(5.32)}$ and $e^{ih(5.83)}$ are small (due to the electrons at these points). Both contribute to $e^{iH (2.43)}$ but \emph{not} to $e^{ih (2.43)}$. This is despite the fact that both $H$ and $h$ are defined \emph{at a point} (namely, at $2.43$). In other words, as far as $h$ is concerned any given point is "listening" to (a part of) its neighborhood, and as far as $H$ is concerned it is "listening" to the points far away, as well. 

The "local" mechanism of performing the above integral is modeled as follows: Every point emits a signal with a certain frequency based on the values of fields $u_1$ and $u_2$. We recall from previous section that $u_1$ and $u_2$ are close to integer most of the time, with an exception of transition regions that are necessary to make sure they are continuous. Their integer approximations correspond to the points in a discrete configuration space (thus, $u_1 (5.12) =2$ and $u_2 (5.12) =1$). Now the frequency of the emitted signals is a function of \emph{both} $u_1$ and $u_2$ (we will call it $\omega (u_1, u_2)$). In light of discreteness, we can define $\omega$ in such a way that $\omega (u_1, u_2) \approx \omega (u_1', u_2')$ if and only if \emph{both} $u_1 \approx u_1'$ \emph{and} $u_2 \approx u_2'$ hold. 

Now, the signals that are to be received with these resonance frequencies communicate the local value of $h$. The \emph{only} other points that can receive that signal are the ones that have the same $u_1$ \emph{and} $u_2$. In other words, they are in the same $S_i \cap T_j$ as the original point. Thus, a signal emitted at $6.37$ can \emph{not} be received at points such as $4.97$, $6. 38$ or $9.73$.  In case of our example, the \emph{only} points that can receive that signal are the ones that fall in the same category based on \emph{both} digits. The examples of the latter are $5.37$, $9.91$, $6.95$, etc. Once these signals are received, they are "converted" into non-oscillating values (based on the mechanism discussed in chapters 5 and 6). As a result the value of $H$ is determined by a sum of the amplitudes of all of the signals that any given point receives. Thus, if we allow the signals propagate much faster than the speed of light, and claim that they can circle a universe within a very short time, then $H$ will approximate the integral of $h$. The details of this will be done in Section 6. 

Another thing that needs to be mentioned is that, despite quasi-locality, $h$ is ultimately expressed \emph{locally} due to our clever way of making particles "look like" they have non-zero size. More precisely, when we say that a particle has a size, we mean that there is some finite differentiable function of space, which represents the density of matter, which happened to be a differentiable approximation to $\delta$-function. Now, in order to accommodate our desired picture, we will continue to view the particles themselves as points, but this density function will be a short-range field that our point particles produce.  Each kind of particle has its own density field: the density field of electron is $\rho_e$, the density field of a proton is $\rho_p$, the density field of positron is $\rho_{e^+}$, etc.  The quasi-local interaction between the particles is due to the \emph{local} interaction of their density fields. 

There is one subtlety here. Namely, \emph{despite} the fact that $h$ "looks" at small but finite neighborhoods, we do \emph{not} want it to look continuous on that scale either. For example, if we consider distances of the order of $0.2$ to be "small" then  $e^{ih(5.12)}$ is large, while $e^{ih(5.13)}$ is small. After all, the "local" electron at $5.32$ is only relevant for the former but not for the latter. In order to account for this, we have to make sure that $\rho$ has similar behavior. So, for example, $\rho_e (7.82) =0$, despite the electron at $7.83$. At the same time,  $\rho_e (7.79)$ is non-zero due to that same electron.  The way the above is accomplished is that each point is "tuned on" to its own frequency. Thus, when a \emph{point} particle occupies a certain location, it sends signals to its neighboring points "telling" them to increase $\rho$. But only the points that are "tuned in" to the same frequency are able to receive these signals. Thus, they are the only ones that actually increase $\rho$. 

Now, in order for the interaction to occur, there has to be two kinds of $\rho$-s: $\rho_1$, which corresponds to the particles that went into the process and $\rho_2$, which corresponds to the particles that are coming out of the process.  Every point has receptors for both of these fields; the latter are set up based on $u_1$ and $u_2$ fields, as usual. The field $\rho_1$ is being received based on $u_1$ which, in our case, corresponds to the second digit after the dot; on the other hand, $\rho_2$ field is being received based on $u_2$ which, in our case, corresponds to the first digit after the dot. Thus, at a point $7.95$, $\rho_2$ is large, while $\rho_1$ is zero. On the other hand, at $7.79$, $\rho_1$ is large while $\rho_2$ is zero. 

In order for $e^{ih}$ to be large, $\rho_1$ and $\rho_2$ have to correlate.  For example, $\rho_1 (5.12)$ is large due to the electron at $5.32$, and $\rho_2 (5.12)$ is large due to the electron at $5.31$. Thus, $e^{ih (5.12)}$ is large. On the other hand, $\rho_1 (5.14)$ is small since electron at $5.31$ no longer counts. On the other hand, $\rho_2 (5.34)$ is still large due to an electron at $5.31$. Thus, $e^{ih (5.14)}$ is small. As one can see, this reproduces the behavior of $e^{ih}$ we have outlined earlier; but this time the mechanism is completely local. 

In case of different kinds of particles, $e^{ih}$ depends on \emph{correlation} of these densities. It closely correlates to Feynman diagrams. For example, since we know that there is a process $e \rightarrow W \nu$, we also know that if $\rho_{1e}$, $\rho_{2W}$ and $\rho_{2\nu}$ are all large, there has to be a \emph{possibility} for $e^{ih}$ to be large as well. The reason I say the word "possibility" is that if there are other particles they might prevent this from being the case. For example, suppose that on top of the above, we also have a proton field $\rho_p$; $\rho_{2p}$ happened to be large while $\rho_{1p}$ is small. Then, the local transition would imply a production of \emph{proton} from weak interaction. Since we do not want the latter, we set $e^{ih}$ to be small \emph{despite} the fact that it would have been large judging from $\rho_e$, $\rho_{\nu}$ and $\rho_W$ alone.  

It is easy to see that the above implies that the dependence of $h$ on $\rho$-fields serves as a replacement of vortexes in Feynman diagrams.  This, of course, requires rewriting quantum field theory in terms of particles that possess size. A lot of it is very non-trivial. For example, Feynman diagrams are generally solved in momentum space; we will now have to redo it for the position space. This might be less obvious than simple Fourier transform since some of the Feynman rules, such as momentum conservation at the vortexes, are designed specifically for momentum space.  Furthermore, in order to allow for propagation of particles, we would have to look at the \emph{derivatives} of the densities. These derivatives will tell us in what direction the particles are displaced from each other, and thus enable us to include that information into a kinetic part of the Hamiltonian. 

Of course, both of the above points call for some machinery that is very different from what is normally being done in quantum field theory. It requires a proposal of a specific function $h(\rho_3, \rho_p, ... )$, followed by a rigorous mathematical proof that this specific function will, indeed, approximate standard quantum field theory (both kinetic and interaction terms).  This is a very difficult project and is far beyond the scope of this paper. For purposes of this paper, we will pretend that this was already done, and show how determinism and locality is restored \emph{once} this is the case. 

\subsection*{4. Desired global picture at the equilibrium}

In the past three sections we have outlined qualitatively what we are trying to accomplish. Now it is time to turn it into more quantitative work. In the next three sections we will quantify the above by means of the local theory of signal propagation. However, since the picture of configuration space is a global one, we would like to go from top to bottom. In this section we will describe the global picture that we would like to obtain at the end of the day. Then, in the subsequent sections, we would attempt to come up with a "local" theory that approximates this global picture at the equilibrium.

Since we would like to use emission and absorption of waves to arrive at the picture, we know that anything that is subject to oscillation changes both magnitude as well as a sign in time. Thus, the only physically meaningful interpretation of $\psi$ is the amplitude of the oscillation. However, the amplitude is always positive and real. In order for $\psi$ to have both imaginary and negative values, we instead define four \emph{distinct} fields $\psi_1$, $\psi_{-1}$, $\psi_i$ and $\psi_{-1}$ and define $\psi$ as 
\beq \psi = \psi_1 - \psi_{-1} + i \psi_i - i \psi_{-i} \eeq
Each of these four fields has \emph{its own} oscillation, and each corresponds to the amplitude of that oscillation. Thus, each of these four fields is positive and real, but the coefficients of $-1$ and $\pm i$ allow $\psi$ to have all possible complex values. One notices that there are many choices of the values of these four fields that give the same value of $\psi$. In particular, $\psi$ is covariant under the transformations
\beq \psi_1 \rightarrow \psi_1 + \chi_r \; ; \; \psi_2 \rightarrow \psi_2 - \chi_r \eeq
and
\beq \psi_i \rightarrow \psi_i + \chi_c \; ; \; \psi_{-i} \rightarrow \psi_{-i} - \chi_c \eeq
For reasons that will become clear later, we will use this freedom and claim that all four fields \emph{grow} in time, and their growth cancels each other in such a way that the resulting field $\psi$ behaves in expected fashion. 
 
Let us now come up with a dynamics of these fields. We will begin by describing dynamics in terms of regular, non-local, configuration space (or, in case of quantum field theory, Fock space), which we will "convert" into our usual space shortly thereafter. To distinguish elements of configuration space (or Fock space) from the ones of usual space, we will denote the former by $S$ and the latter by $\vec{x}$ (notice that only the elements of the regular space have a vector sign). We will assume that the configuration space is discrete. Therefore, the most general dynamics for $\psi$ in a configuration space is given by 
\beq \frac{\partial (\psi (t, S_k))}{\partial t} = \sum_l i H (S_k, S_l) \psi (x_l) \eeq
Just like we did with $\psi$, we will break $H$ into four pieces: 
\beq H = H_1 - H_{-1} + i H_i - i H_{-i}\eeq
By substituting the values of $\psi$ and $H$, the right hand side of the evolution equation becomes
\beq \sum_l i H(S_k, S_l) \psi (S_l) = \sum_l (H_1 (S_k, S_l) \psi_1 (S_l) + H_{-1} (S_k, S_l) \psi_{-1} (S_l) \nonumber \eeq
\beq +H_i (S_k, S_l) \psi_{-i} (S_l) + H_{-i} (S_k, S_l) \psi_i (S_l) -H_1 (S_k, S_l) \psi_{-1} (S_l) -\nonumber \eeq
\beq  - H_{-1} (S_k, S_l) \psi_1 (S_l) -H_i (S_k, S_l) \psi_i (S_l) - H_{-i} (S_k, S_l) \psi_{-i} (S_l) + \eeq
\beq + i H_1 (S_k, S_l) \psi_i (S_l) + i H_{-1} (S_k, S_l) \psi_{-i} (S_l) + i H_i (S_k, S_l) \psi_1 (S_l) +  \nonumber \eeq
\beq + i H_{-i} (S_k, S_l) \psi_{-1} (S_l)  - i H_1 (S_k, S_l) \psi_{-i} (S_l) -iH_{-1} (S_k, S_l) \psi_i (S_l) - \nonumber \eeq
\beq - iH_i(S_k, S_l) \psi_{-1} (S_l)  - iH_{-i} (S_k, S_l) \psi_1 (S_l) \nonumber\eeq

By equating it to $\partial \psi (t, S_k) / \partial t$, and remembering that each of the $\psi_1$, $\psi_{-1}$, $\psi_i$ and $\psi_{-i}$ has to be positive, we read off the dynamic equations for these four fields:
\beq \partial_t \psi_1 (S_k, t) = \sum_l (H_1 (S_k, S_l) \psi_{-i} (S_l, t) + H_{-1} (S_k, S_l) \psi_i (S_l, t) + \nonumber \eeq
\beq +  H_i (S_k, S_l) \psi_{-1} (S_l, t)) +H_{-i} (S_k, S_l) \psi_1 (S_l, t) \nonumber \eeq
\beq \partial_t \psi_{-1} (S_k, t) = \sum_l (H_1 (S_k, S_l) \psi_i (S_l, t) + H_{-1} (S_k, S_l) \psi_{-i} (S_l, t)+ \nonumber \eeq
\beq + H_i (S_k, S_l) \psi_1 (S_l, t)+ H_{-i} (S_k, S_l) \psi_{-1} (S_l, t)) \eeq
\beq \partial_t \psi_i (S_k, t) = \sum_l (H_1 (S_k, S_l) \psi_1 (S_l, t) + H_{-1} (S_k, S_l) \psi_{-1} (S_l, t)+ \nonumber \eeq
\beq + H_i (S_k, S_l) \psi_{-i} (S_l, t)) + H_{-i} (S_k, S_l) \psi_i (S_l, t) \nonumber \eeq
\beq \partial_t \psi_{-i} (S_k, t) = \sum_l (H_1 (S_k, S_l) \psi_{-1} (S_l, t) + H_{-1} (S_k, S_l) \psi_1 (S_l, t) + \nonumber \eeq
\beq + H_i (S_k, S_l) \psi_{i} (S_l, t)) + H_{-i} (S_k, S_l) \psi_{-i} (S_l, t) \nonumber \eeq
We would now like to describe a corresponding dynamics in the usual space. We recall that $S_k$ is both a point in a configuration space as well as a subset of a regular space; namely, a set of points $\vec{x}$ satisfying $u_1 (\vec{x}) \approx k$. To make this more precise, we can define $S_k$ as 
\beq S_k = \{ \vec{x} \in \mathbb{R}^3 \vert \vert u_1 (\vec{x}) - k \vert < \epsilon \} \eeq
for some small $\epsilon$. In light of the fact that $u_1 (x)$ (just like $u_2 (x)$) can be approximated by integers for almost all $x$, the union of all $S_k$ covers most of our space. The small "gaps" that are left as a result of continuity of $u_1$ are not very important. The value of $\psi (S_k)$ is represented by $\psi (\vec{x})$ for \emph{any} $x \in S_k$. Therefore, whenever $u_1 (\vec{x})$ and $u_1 (\vec{x}')$ are both approximated by the same integer $k$, we would like to have 
\beq \psi_1 (\vec{x}) \approx \psi_1 (\vec{x}') \; ; \; \psi_{-1} (\vec{x}) \approx \psi_{-1} (\vec{x}') \; ; \; \psi_i (\vec{x}) \approx \psi_i (\vec{x}') \; ; \; \psi_{-i} (\vec{x}) \approx \psi_{-i} (\vec{x}')  \eeq
The fields $H_1$, $H_{-1}$, $H_i$ and $H_{-i}$ should also be translated into the usual spacetime. By inspection, $H (S_k, S_l)$ is defined to be a function of \emph{two} points in the configuration space: the one that emits a signal ($S_l$) and the one that receives it ($S_k$). Thus, we should identify a region in a regular space, $U_{kl}$, that somehow "encodes" both $S_l$ and $S_k$, and $H$ should be nearly constant within that region. That is the main purpose why two fields, $u_1$ and $u_2$ were introduced. In particular, we define it to be $S_k \cap T_l$,
 where 
\beq T_l = \{ \vec{x} \vert   u_2 (\vec{x}) - l \vert < \epsilon \} \eeq
This means that, in order for $H_1$, $H_{-1}$, $H_i$ and $H_{-i}$ to be consistently defined, we would like to introduce the dynamics of these four fields in such a way that 
\beq H_1 (\vec{x}) \approx H_1 (\vec{x}') \; ;\; H_{-1} (\vec{x}) \approx H_{-1} (\vec{x}') \; ; \; H_i (\vec{x}) \approx H_i (\vec{x}') \; ; \; H_{-i} (\vec{x}) \approx H_{-i} (\vec{x}')\eeq
whenever \emph{both} $u_1 (\vec{x}) \approx u_1 (\vec{x'})$ \emph{and} $u_2 (\vec{x}) \approx u_2 (\vec{x'})$ hold. Now, once we have established the consistency of $\psi$-s and $H$-s, the "translation" of the four evolution equations becomes 
\beq \partial_t \psi_1 (\vec{x}, t) \approx \int_{T_l} d^3 x (H_1 (\vec{x}') \psi_{-i} (S_l, t) + H_{-1}  (\vec{x}') \psi_i (S_l, t) + \nonumber \eeq
\beq + H_i (\vec{x}') \psi_{-1} (\vec{x}', t)) + H_{-i} (\vec{x}') \psi_{-1} (S_l, t)) \nonumber \eeq
\beq \partial_t \psi_{-1} (\vec{x}, t) \approx \int_{T_l} d^3 x  (H_1 (\vec{x}') \psi_i (S_l, t) + H_{-1} (\vec{x}') \psi_{-i} (S_l, t) + \nonumber \eeq
\beq +  H_i (\vec{x}') \psi_1 (\vec{x}', t) + H_{-i} (\vec{x}') \psi_{-1} (S_l, t) ) \eeq
\beq \partial_t \psi_i (\vec{x}, t) \approx  \int_{T_l} d^3 x (H_1 (\vec{x}') \psi_1 (S_l, t) + H_{-1} (\vec{x}') \psi_{-1} (S_l, t) + \nonumber \eeq
\beq + H_i (\vec{x}') \psi_{-i} (\vec{x}', t) + H_{-i} (\vec{x}') \psi_i (S_l, t) ) \eeq
\beq \partial_t \psi_{-i} (\vec{x}, t) \approx \int_{T_l} d^3 x (H_1 (\vec{x}') \psi_{-1} (S_l, t) + H_{-1} (\vec{x}') \psi_1 (S_l, t) + \nonumber \eeq
\beq + H_i (\vec{x}') \psi_{i} (\vec{x}', t) + H_{-i} (\vec{x}') \psi_{-i} (S_l, t) ) \nonumber \eeq
To see why it works, consider the first equation. Since there is a one to one correspondence between $S_k \cap T_l$ and $(u_1, u_2)$, we know that different $S_k \cap T_l$ fill most of the space. Thus the integral can be approximated as a sum of integrals over each $S_k \cap T_l$. Now, in the interior of $S_k \cap T_l$, $H_i (\vec{x}) \approx H(S_k, S_l)$ and $\psi_{-i} (\vec{x}, t) \approx \psi_{-i} (S_l, t)$. Therefore, if the volume of $S_k \cap T_l$ is denoted by $V(S_k \cap T_l)$, the expression becomes
\beq \partial_t \psi_1 (\vec{x}', t) \approx \sum_{k,l} V(S_k \cap T_l)(H_1  (S_k, S_l) \psi_{-i} (S_l, t) + H_{-1}  (S_k, S_l) \psi_{i} (S_l, t) + \nonumber \eeq
\beq + H_i (S_k, S_l) \psi_{-1} (S_l, t) + H_{-i}  (S_k, S_l) \psi_1 (S_l, t) ) \nonumber \eeq
Now, due to a very large volume of the universe as well as very fast fluctuations of $u_1$ and $u_2$, the volumes of different $S_k \cap T_l$ approximate a common value $V$. Thus, the expression becomes
\beq \partial_t \psi_1 (\vec{x}', t) \approx V \sum_{k,l} (H_1  (S_k, S_l) \psi_{-i} (S_l, t) + H_{-1}  (S_k, S_l) \psi_{i} (S_l, t) + \nonumber \eeq
\beq + H_i (S_k, S_l) \psi_{-1} (S_l, t) + H_{-i}  (S_k, S_l) \psi_1 (S_l, t) ) \nonumber \eeq
which is the same as the original expression we had for configuration space. The other three equations work out similarly. 

Therefore, our goal for the sections that follow can be summarized as follows:

1) Describe dynamics of $\psi$ in such a way that

a) $\psi$ is nearly the same at points with similar values of $u_1$

b) $\psi$ can be communicated between different points by the integral equations given above

2) Describe dynamics of $H$ in such a way that

a) $H$ is nearly the same at $\vec{x}$ and $\vec{x'}$ as long as $u_1 (\vec{x}) \approx u_1 (\vec{x}')$ \emph{and} $u_2 (\vec{x}) \approx u_2 (\vec{x}')$

b) $H$ corresponds to the \emph{global} Hamiltonian predicted by standard quantum mechanics (or standard quantum field theory). 

In the next section we will assume that $H$ is already given, and we will focus on dynamics of $\psi$. Then, in the section after that, we will return to $H$.

\subsection*{5. Dynamics of $\psi$ when $H$ is given}

As was mentioned before, in terms of our usual space, the points $S_k$ and $S_l$ in configuration space are defined as 
\beq S_k = \{ \vec{x} \in \mathbb{R}^3 \vert \; \vert u_1 ( \vec{x}) - k \vert < \epsilon \} \; ; \; S_l = \{ \vec{x} \in \mathbb{R}^3 \vert \; \vert u_1 ( \vec{x}) - l \vert < \epsilon \} \eeq
for some small $\epsilon$. Our desired dynamical equation represents the communication from $S_l$ to $S_k$. That communication happens by means of points $\vec{x}$ simultaneously satisfying $u_1 (\vec{x}) \approx l$ and $u_2 (\vec{x}) \approx k$. The signal emitted by these points is to be received by points $\vec{x'}$ satisfying $u_1 (\vec{x'}) \approx k$. Thus, we would like the frequency of emitted signal to be approximately $k$. This can be accomplished by postulating a dynamics that guarantees frequency of emitted signal to be $u_2 (\vec{x})$. 

We then encounter a problem: as the signal travels through space, the $u_2$ changes, which means that we run the risk of the frequency of that signal changing as well. We will simply avoid this issue by coming up with a mechanism of random "sparks" that trigger the emission of signals with frequency $u_2$. Since these sparks are localized both in space and in time, only the value of $u_2$ at the place of their occurrence is important. After the signal had been produced, its subsequent motion is identical to the one in a vacuum, regardless of the behavior of $u_2$. Thus, original frequency, $u_2 (\vec{x}_0)$ is preserved.

The specific mechanism of spark production is not very important. For definiteness, we postulate some real valued field $\chi$ that evolves according to usual wave equation, 
\beq \frac{\partial^2 \chi}{\partial t^2} - c_{\chi} \vec{\nabla}^2 \chi =0 \eeq
and simply postulate that the intensity of a "spark" is given by 
\beq f(\chi) = \frac{1}{2} + \frac{1}{\pi} tan^{-1} (a_{\chi} (\chi - \chi_0)) \eeq
where $a_{\chi}$ is a very large number. Thus, $f(\chi)$ is close to $0$ if $\chi < \chi_0 - \delta$, and it is close to $1$ if it is greater than $\chi_0 + \delta$, for some small $\delta$. Due to the compactness of the universe, sometimes different Fourier components of $\chi$ will produce local resonances. $\chi_0$ is chosen in such a way that it is not likely for $\chi$ to exceed $\chi_0$; thus $f(\chi)$ looks like a sequence of random sparks. At the same time, $\chi_0$ is small enough for these "sparks" not to be too rare. 

Now, we are ready to discuss the emission of the signal. The sources of emission of $\psi_1$, $\psi_{-1}$, $\psi_i$ and $\psi_{-1}$ will be denoted as $e_{\psi_1}$, $e_{\psi_{-1}}$, $e_{\psi_i}$ and $e_{\psi_{-1}}$, respectively. In order to obtain our desired equations, we would like the amplitude of these pulses to be determined by the couplings of $f(\chi)$, $H$-s, and $\psi$-s, where the indexes in the two latter fields correspond to the indexes in our desired equations. Since we want $e$-s to be localized in space, we do \emph{not} include spatial derivatives in their dynamics, thus making sure that they don't propagate. Thus, we propose the following equations:
\beq \partial_0^2 e_{\psi_1}(t, x) = - u_2 (\vec{x}) e_{\psi_1} (t, \vec{x}) - \lambda_e \partial_0 e_{\psi_1} (t, \vec{x})  +   \nonumber \eeq
\beq + f(\chi (t, \vec{x})) (H_1 (\vec{x}) \psi_{-i} (t, \vec{x}) + H_{-1} (\vec{x}) \psi_i (t, \vec{x}) + \nonumber \eeq
\beq + H_i (\vec{x}) \psi_{-1} (t, \vec{x}) + H_{-i} (\vec{x}) \psi_1 (t, \vec{x})) \nonumber \eeq
\beq \partial_0^2 e_{\psi_{-1}}(t, x) = - u_2 (\vec{x}) e_{\psi_{-1}} (t, \vec{x}) - \lambda_e \partial_0 e_{\psi_{-1}} (t, \vec{x}) + \nonumber \eeq
\beq + f(\chi (t, \vec{x})) (H_1 (\vec{x}) \psi_i (t, \vec{x})+  H_{-1} (\vec{x}) \psi_{-i} (t, \vec{x}) + \nonumber \eeq
\beq +  H_i (\vec{x}) \psi_1 (t, \vec{x})+ H_{-i} (\vec{x}) \psi_{-1} (t, \vec{x}) \nonumber \eeq
\beq \partial_0^2 e_{\psi_i}(t, x) = - u_2 (\vec{x}) e_{\psi_i} (t, \vec{x}) - \lambda_e \partial_0 e_{\psi_i} (t, \vec{x}) + \eeq
\beq +  f(\chi (t, \vec{x}))( H_1 (\vec{x}) (\psi_1 (t, \vec{x}) +  H_{-1} (\vec{x}) \psi_{-1} (t, \vec{x})+ \nonumber \eeq
\beq +  H_i (\vec{x}) \psi_{-i} (t, \vec{x}) + H_{-i} (\vec{x}) \psi_{i} (t, \vec{x}) \nonumber \eeq
\beq \partial_0^2 e_{\psi_{-i}}(t, x) = - u_2 (\vec{x}) e_{\psi_{-i}} (t, \vec{x}) - \lambda_e \partial_0 e_{\psi_{-i}} (t, \vec{x}) + \nonumber \eeq
\beq +  f(\chi (t, \vec{x}))( H_1 (\vec{x}) \psi_{-1} (t,  \vec{x})  + H_{-1} (\vec{x}) \psi_1 (t, \vec{x}) + \nonumber \eeq
\beq +  H_i (\vec{x}) \psi_i (t, \vec{x}) + H_{-i} (\vec{x}) \psi_{-i} (t, \vec{x} )\nonumber \eeq
In the above equations, the first two terms assure that whenever an excitation occurs, a short pulse is sent, with frequency $u_2$. The last two terms tell us that the source of excitation is a coupling of $f(\chi)$, $H$ and $\psi$.  The indexes selected in a way that correspond to our desired evolution equations of four $\psi$ fields, written earlier.  As we have pointed out earlier, the above equations don't have space derivatives, and thus $e$-s do not propagate. The communication between different points in space is enforced through \emph{messenger fields}, $\mu_{\psi_1}$, $\mu_{\psi_{-1}}$, $\mu_{\psi_i}$ and $\mu_{\psi_{-i}}$. Their dynamics is given by
\beq \partial_t^2 \mu_{\psi_1} - c_{\mu}^2 \nabla^2 \mu_{\psi_1} = e_{\psi_1} \; ; \; \partial_t^2 \mu_{\psi_{-1}} - c_{\mu}^2 \nabla^2 \mu_{\psi_{-1}} = e_{\psi_{-1}} \nonumber \eeq
\beq \partial_t^2 \mu_{\psi_i} - c_{\mu}^2 \nabla^2 \mu_{\psi_i} = e_{\psi_i} \; ; \; \partial_t^2 \mu_{\psi_{-i}} - c_{\mu}^2 \nabla^2 \mu_{\psi_{-i}} = e_{\psi_{-i}}  \eeq
where $\nabla^2$ represents spacial  Laplassian and $c_{\mu}$ is some very large constant that corresponds to the speed of propagation of $\mu$. Once again, none of the speeds of the signals have anything in common with the speed of light; in fact, they are expected to be much larger than that. Thus, we have found a mechanism by which the signals are emitted with frequency $u_2 (\vec{x}_0)$, as desired. Since their further propagation is mathematically identical to the one in a vacuum, their frequency stays constant. Thus, when that signal passes a point $\vec{x}$ for which $u_2(\vec{x}) \neq u_2(\vec{x}_0)$, its frequency continues to be equal $u_2 (\vec{x}_0)$. 

Finally, these fields are received by the corresponding "receptors" at the desired locations, $r_{\psi_1}$, $r_{\psi_{-1}}$, $r_{\psi_i}$ and $r_{\psi_{-i}}$. As mentioned earlier, the absorption resonance frequency is based on $u_1$ rather than $u_2$ and the process is defined as follows:
\beq \partial_t^2 r_{\psi_1} (\vec{x}, t) = - u_1 (\vec{x}) r_{\psi_1} (t, \vec{x}) - \lambda_r \partial_0 r_{\psi_1} (t, x) + \mu_{\psi_1} (t, x) \nonumber \eeq
\beq \partial_t^2 r_{\psi_{-1}} (\vec{x}, t) = - u_1 (\vec{x}) r_{\psi_{-1}} (t, \vec{x}) - \lambda_r \partial_0 r_{\psi_{-1}} (t, x) + \mu_{\psi_{-1}} (t, x) \nonumber \eeq
\beq \partial_t^2 r_{\psi_i} (\vec{x}, t) = - u_1 (\vec{x}) r_{\psi_i} (t, \vec{x}) - \lambda_r \partial_0 r_{\psi_i} (t, x) + \mu_{\psi_i} (t, x) \eeq
\beq \partial_t^2 r_{\psi_{-i}} (\vec{x}, t) = - u_1 (\vec{x}) r_{\psi_{-i}} (t, \vec{x}) - \lambda_r \partial_0 r_{\psi_{-i}} (t, x) + \mu_{\psi_{-i}} (t, x) \nonumber \eeq
The above equations are mathematically identical to the driven harmonic oscillator in $r_{\psi}$, with sinusoidal driving force $\mu_{\psi}$. While $\lambda_e$ was assumed to be large enough for the $e$-oscillation to have a short lifetime, we make the opposite assumption about $\lambda_r$: it is assumed to be very small. As a result, whenever $u_1 (\vec{x}')$ is close to the frequency of $\mu_{\psi}$, this results in a very large resonance in $r_{\psi}$. But, the frequency of $\mu_{\psi}$ is equal to $u_2 (x)$, where $\vec{x}$ is a point of emission of a signal. Thus, whenever $u_2 (\vec{x}) \approx u_1 (\vec{x}')$, $\vec{x}'$ absorbs signals emitted at $\vec{x}$. On the other hand, if $u (\vec{x})$ and $u (\vec{x}')$ are different, the fact that both are approximated by integers implies that their difference has to be an integer; that difference is large enough for the signal to go virtually unnoticed.

We now assume that our space is compact (say, a large sphere). Then the emitted signals circle our universe multiple times. Noticeably, the "absorption" of $\mu_{\psi}$ in above dynamics is not conservative: it \emph{only} has an effect on $\psi$, but \emph{not} on $\mu_{\psi}$. The emission, on the other hand, \emph{does} have an effect on $\mu_{\psi}$. As a result, $\mu_{\psi}$ fields accumulate over time.  Most of them are "old enough" to have circled our universe multiple times, which means that location of their emission is no longer important: they are just as intense around the points of their emission as they are billion light years away. 

If their speed is large enough for them to circle the universe within very short time, then, on larger time scales, the $\mu$-signals emitted during a physical process have global effect, which is independent of location.  This is one of the key features that leads to the observed non-locality of the configuration space. Noticeably, their accumulation in time leads to the increase of $\psi_1$, $\psi_{-1}$, $\psi_i$ and $\psi_{-i}$. But, as mentioned earlier, when it comes to the total field $\psi$, the increases of these fields pairwise cancel, which results in the expected behavior of $\psi$. 

We know from the theory of oscillations that the \emph{squares} of the amplitudes of out-of-phase oscillations add up. So, since we have established that the amplitude of our oscillations is position-independent, we will get the desired dynamics of $\psi_1$, $\psi_{-1}$, $\psi_i$ and $\psi_{-i}$ if we associate them with \emph{squares} of amplitudes of oscillation of $r_{\psi_1}$, $r_{\psi_{-1}}$, $r_{\psi_i}$ and $r_{\psi_{-i}}$, respectively. Since these are four separate fields, each of these four relations is enforced separately, independent of the other three. In other words, contrary to what we are used to, each of the "components" of $\psi$ is, itself, a square of the amplitude.  

This can be accomplished by setting up the dynamics of $\psi_1$ in such a way that whenever $\psi_1$ is less than $r_{\psi_1}^2$, it "catches up" \emph{fast}; but if it is greater than $r_{\psi_1}^2$, it decreases \emph{slowly}. Suppose the peak of the oscillation of $r_{\psi_1}$ is $R_{\psi_1}$. Then, as soon as $r_{\psi_1}^2$ has reached $R_{\psi_1}^2$, the $\psi_1$ \emph{quickly} reaches the same. But then when $r_{\psi_1}^2$ goes back to $0$, the $\psi_1$ "doesn't have time" to decrease much, until $r_{\psi_1}^2$ is back to $R_{\psi_1}^2$ again. Thus, if the frequency of oscillations is high enough, $\psi_1 \approx R_{\psi_1}^2$ \emph{at all times}. 

On the other hand, if the amplitude of the oscillations has changed from $R_{\psi_1}$ to $S_{\psi_1}< R_{\psi_1}$, then after enough time passes, the $\psi_1$ field will finally reach $S_{\psi_1}^2$ (even though it might take several oscillations for this to occur). Thus, the amplitude of $\psi_1$ will behave the way it is expected on a larger time scales. The same, of course, is true for $\psi_{-1}$, $\psi_i$ and $\psi_{-i}$. Thus, we postulate the dynamics of these four fields to be 
\beq \partial_0^2 \psi_1 = a (e^{b(r_{\psi_1}^2 - \psi_1)} -1) \; ; \; \partial_0^2 \psi_{-1} = a (e^{b(r_{\psi_{-1}}^2 - \psi_{-1})} -1) \nonumber \eeq
\beq \partial_0^2 \psi_i = a (e^{b(r_{\psi_i}^2 - \psi_i)} -1) \; ; \; \partial_0^2 \psi_{-i} = a (e^{b(r_{\psi_{-i}}^2 - \psi_{-i})} -1)  \eeq
It is easy to see that if $a$ is very small and $b$ is very large, this will lead to the desired result. 

 We have just established that the field $\psi_1$ is proportional to the square of the amplitude of oscillation of $r_{\psi_1}$. Furthermore $r_{\psi_1}$ is simply a driven harmonic oscillator with a driving force $\mu_{\psi_1}$.Thus, from the theory of driven harmonic oscillator, we know that the amplitude of oscillations of $r_{\psi_1}$ is proportional to the amplitude of $\mu_{\psi_1}$. By inspection our desired equation from the previous chapter,
\beq \partial_t \psi_1 (\vec{x}', t) \approx \int d^3 x (H_1 (\vec{x}) \psi_{-i} (S_l, t) + H_{-1} (\vec{x}) \psi_i (S_l, t) + \nonumber \eeq
\beq + H_i (\vec{x}) \psi_{-1} (\vec{x}, t)) + H_{-i} (\vec{x}) \psi_1 (S_l, t) ,\eeq
the latter is made up by incoherent signals coming from the collisions of $(\psi, -i)$ with  $(H, r)$ and $(\psi, -1)$ with $(H, i)$.

\subsection*{6. Dynamics for H}

In the previous section we have assumed that $H$ was only a function of $\vec{x}$ and \emph{not} of $t$. Physically, this means that the probability amplitude of transition between any two given states is the same, at all times. The only thing that changes with time is their probability amplitudes. However, while this agrees with the standard view, we are not happy with that view. After all, we don't like the fact that transition probabilities between different configurations are non-local. Therefore, we would like to come up with a \emph{local} mechanism by which $H$ attains this global value. We have shown a jist of it in section 3, when we have discussed the difference between $h$ and $H$.

The good news is that the sought-after mechanism should be based on the configuration spaces \emph{alone}; that is, the configuration of the particles. As was explained in previous sections, particles are stationary. Therefore, the task of obtaining global equilibrium for $H$ is much easier than the one for $\psi$: $H$ can evolve as slowly as we wish, since it doesn't have to "keep up" with any changes. The challenge, however, is that, at any given point $\vec{x}$, with $u_1 (\vec{x}) \approx k$ and $u_2 (\vec{x}) \approx l$, we would like $H (x)$ to look \emph{only} at the states $\vert s_1 >$ and $\vert s_2 >$. 

In order to be able to "filter out" the rest of the information, we will have to introduce harmonic processes with frequencies $u_1$ and $u_2$ at every point. Formally, this can be done by introducing two fields $\alpha_1$ and $\alpha_2$. In order not to "confuse" the oscillatory processes associated with different points, we should make sure these fields do \emph{not} propagate. Thus, their dynamics does not have space derivatives. We postulate their equations of motion to be 
\beq \partial_t^2 \alpha_1 (\vec{x}, t)  = - u_1^2 (\vec{x}) \alpha_1 (\vec{x}, t) \; ; \; \partial_t^2 \alpha_2 (\vec{x}, t) = - u_2^2 (\vec{x}) \alpha_2 (\vec{x}, t) \eeq
The above equations, however, allow for solution to alter in space both its phase and amplitude:
\beq \alpha_1 (\vec{x}, t) = A_1 (\vec{x}) \; cos (t - t_1 (\vec{x})) \; ; \; \alpha_2 (\vec{x}, t) = A_2 (\vec{x}) \; cos (t - t_2 (\vec{x})). \eeq
Since we do not want the above variations in amplitude, we introduce another two fields, $\beta_1$ and $\beta_2$, which are the normalizations of $\alpha_1$ and $\alpha_2$ respectively:
\beq \beta_1 (x, t) = \frac{u_1 (x) \alpha_1 (x, t)}{\sqrt{u_1^2 (x) \alpha_1^2 (x, t) + (\partial_t \alpha_1)^2 (x, t)}} \; ; \; \beta_2 (x, t) = \frac{u_2 (x) \alpha_2 (x, t)}{\sqrt{u_2^2 (x) \alpha_2^2 (x, t) + (\partial_t \alpha_2)^2 (x, t)}}. \eeq 
Of course, the phases $t_1 (\vec{x})$ and $t_2 (\vec{x})$ can not be dealt with this way; but as we shall see, they will eventually become unimportant. 

Now, in order to introduce the Hamiltonian, we have to first come up with "local" way of defining it point-wise and then "integrating" it. Since particles have zero size, they don't "touch" each other. Thus, we do not have literal vortexes. Instead, we need to define a "density field" associated with each particle that forms a cloud around it. While the particle is still a point object, that density field is extended in space. Thus, the overlap of the density fields of, say, electron, neutrino, and W-boson, forms $e \nu W$ vortex. Of course, this calls for rewriting quantum field theory in a more continuous way. This is far beyond the scope of this paper.

As far as this paper is concerned. we will assume that we do have a version of quantum field theory that allows particles to have size, and our only goal is reinterpreting phase space and introducing determinism. However, there is one important thing to address: if a particle has size, it will undoubtfully overlap with different $S$-s. But we want that particle to be detected \emph{only} in the $S _{u_1 (\vec{x_0})}$ and $S_{u_2 (\vec{x}_0)}$, where $x_0$ is a location of a "center" of the particle. We will do that by introducing the emission and reception mechanism of $\rho$-field. That field is being "emitted" at $\vec{x}_0$ and then "received" at $\vec{x}$ that belongs to the same $S_u$. 

Let us now be a little bit more explicit. Suppose we have $A$ types of particles. For example, particle number $1$ is electron, particle number $2$ is proton, etc. We will denote the density field corresponding to particle number $A$ by $\rho_A$. The "source" of that $\rho$ field is a "point charge" $q_A$, and its "messenger" is $\mu_{\rho_A}$. The latter undergoes oscillations, which are guided by the behavior of $\beta$ at the center:
\beq \mu_{\rho_{A}} (\vec{x}) = \sum_k \frac{q_A}{\vert \vec{x} - \vec{x}_k \vert} \beta \Big(\vec{x}_k,  t - \frac{\vert \vec{x} - \vec{x}_k \vert}{c_{\rho}} \Big), \eeq
where $k$ is the numbering of different particles of the same type $A$ (they are assumed to be distinguishable), and $\vec{x}_{Ak}$ is their location in space (there is no $t$-dependence of $\vec{x}_{Ak}$ because, as explained earlier, we have assumed they are stationary). in the above expression time delay is introduced for the purposes of locality. The inverse dependence on distance assures a very small size of density distribution, given the appropriate value of $q_A$. Since any resemblance with electrostatics is merely the issue of convenience, any differences between the two theories are not very important. 

Now, at any given point, we would like to define a local process. For example, suppose near a given point there is an electron, a neutrino and $W$ boson. We would like to say, for example, that what contributed to Hamiltonian is the electron emitting $W$ boson, and turning into neutrino. Thus, we would like to view electron as part of $S_{u_1 (\vec{x})}$, while viewing both neutrino and $W$ boson as parts of $S_{u_2 (\vec{x})}$. This can be accomplished if electron, neutrino and $W$-boson are placed at points $\vec{x}_1$, $\vec{x}_2$ and $\vec{x}_3$, all of which are close to $\vec{x}_0$ and which satisfy $u_1 (\vec{x}_1) = u_1 (\vec{x}_0)$, $u_2 (\vec{x}_2) = u_2 (\vec{x}_0)$ and $u_2 (\vec{x}_3) = u_2 (\vec{x}_0)$. Thus, a point $\vec{x}_0$ needs to have receptors for both $u_1$ and $u_2$. The same applies to every other point. Thus, we define two separate receptors:
\beq \frac{\partial r_{\rho_{A1}}}{\partial t^2} = \mu_{\rho_A} (\vec{x}, t) - u_1^2 (\vec{x}) r_{\rho 1} (\vec{x}, t) - \lambda_{r_{\rho_{A1}}} \frac{\partial r_{\rho_{A1}}}{\partial t} \eeq
\beq \frac{\partial r_{\rho_{A2}}}{\partial t^2} = \mu_{\rho_A} (\vec{x}, t) - u_2^2 (\vec{x}) r_{\rho 2} (\vec{x}, t) - \lambda_{r_{\rho_{A2}}} \frac{\partial r_{\rho_{A2}}}{\partial t} \eeq
As before, the receptor fields oscillate; but we do not want our $\rho$ to do that. We use the same tactic as in previous section, and "convert" receptor fields into non-oscillating densities via the following expressions:
\beq \frac{\partial^2 \rho_{A1}}{\partial t^2} = a_{\rho} (e^{b_{\rho} (r_{\rho_{A1}}^2 - \rho_{A1})} -1) \; ; \; \frac{\partial^2 \rho_{A2}}{\partial t^2} = a_{\rho} (e^{b_{\rho} (r_{\rho_{A2}}^2 - \rho_{A2})}  -1)\eeq 
We are now ready to move on to define Hamiltonian. As stated before, we are dealing with modified version of quantum field theory (which we never introduced but we pretend that we did) that is based on \emph{densities} of the particles as opposed to particles themselves. Thus, if the above local neighborhood was all there is in the universe we would simply say that H is a function of $\rho$-s. But since that is not the case, we would have to "integrate". The way to do that locally is to follow previous procedure and introduce emitter ($e_H$), messenger field ($\mu_H$) and receiver field ($r_H$). Thus, only the emitter field is the local function of above described densities.

However, given that our emitters are placed fixed distances away from each other, if the emission was constant that would lead to interference. Since we do not want that, we would like to replace a continuous process with discrete emissions. When emissions do occur, they have prescribed frequencies; but, at the same time, they are pulse-like. If we make sure that they are random in time, the total phase shift between emissions at any two given points will be random, despite the fact that the distance is known. This will assure us that, on average, the squares of amplitudes add up, regardless of the distance.

In order to come up with \emph{deterministic} mechanism of producing "random pulses" we need a "random generator". For that purpose, we introduce another field, $\chi (x, t)$, that evolves according to usual wave equation,
\beq \partial_t^2 \chi - c_{\chi}^2 \nabla^2 \chi = 0. \eeq
As with all other fields, $c_{\chi}$ is much larger than the speed of light. In fact it is so large that within small time interval, the field $\chi$ circles our universe multiple times (as stated before, we assume that our universe is compact). Since the exact shape of the universe is not known, there are might be some resonances of $\chi$ at random places. We now introduce a differentiable approximation to step function,
\beq f(\theta) = \frac{1}{2} + tan^{-1} \; (\theta - \theta_0) \eeq
If we choose $\theta_0$ in such a way that $\chi(t, \vec{x})$ exceeds $\theta_0$ only at very unlikely resonances, it will mean that  $f(\chi (t, \vec{x}))$ is $1$ at these rare instances and $0$ everywhere else. In other words, the latter will look like a set of randomly distributed pulses, which is what we need. Thus, we define our emission equations for $H_1$, $H_{-1}$, $H_i$ and $H_{-i}$ as follows:
\beq \partial_t^2 e_{H_1} (x, t) = - \omega^2 (u_1 (x), u_2 (x)) e_{H_1} (\vec{x}, t) - \lambda_e \partial_0 e_{H_1} (\vec{x}, t) + \nonumber \eeq
\beq + f(\chi (\vec{x}, t)) h_1(\rho_{11} (x, t) , . . . , \rho_{B1} (x,t); \rho_{12} (x, t) , . . . , \rho_{B2} (x,t)) \nonumber \eeq
\beq \partial_t^2 e_{H_{-1}} (x, t) = - \omega^2 (u_1 (x), u_2 (x)) e_{H_{-1}} (\vec{x}, t) - \lambda_e \partial_0 e_{H_{-1}} (\vec{x}, t) + \nonumber \eeq
\beq + f(\chi (\vec{x}, t)) h_{-1}(\rho_{11} (x, t) , . . . , \rho_{B1} (x,t); \rho_{12} (x, t) , . . . , \rho_{B2} (x,t)) \eeq
\beq \partial_t^2 e_{H_i} (x, t) = - \omega^2 (u_1 (x), u_2 (x)) e_{H_i} (\vec{x}, t) - \lambda_e \partial_0 e_{H_i} (\vec{x}, t) + \nonumber \eeq
\beq + f(\chi (\vec{x}, t)) h_i(\rho_{11} (x, t) , . . . , \rho_{B1} (x,t); \rho_{12} (x, t) , . . . , \rho_{B2} (x,t)) \nonumber \eeq
\beq \partial_t^2 e_{H_{-i}} (x, t) = - \omega^2 (u_1 (x), u_2 (x)) e_{H_{-i}} (\vec{x}, t) - \lambda_e \partial_0 e_{H_{-i}} (\vec{x}, t) + \nonumber \eeq
\beq + f(\chi (\vec{x}, t)) h_{-i}(\rho_{11} (x, t) , . . . , \rho_{B1} (x,t); \rho_{12} (x, t) , . . . , \rho_{B2} (x,t)) \nonumber \eeq
In the above expression it doesn't matter that $e_{H_r}$ and $e_{H_i}$  are simultaneous both in space and time. After all, from the previous chapter we remember that we did \emph{not} directly add $\psi_1$, $\psi_{-1}$, $\psi_i$ and $\psi_{-i}$ until the very end, at which point we no longer had to worry about their oscillations. Similarly, we do not add $H_r$ and $H_{-1}$ either until that point. Thus, the only thing we need to worry about is the interference of these fields \emph{with themselves}, and \emph{not} with each other.  

Since $f(\chi (\vec{x}, t))$ is either $0$ or $1$, the amplitude of the pulses are determined by $h$. Thus, losely speaking, after the propagation, $h^2$ will globally "add" to give us $H$. The mechanism of this propagation is similar to the one of $\psi$. We introduce the "messenger fields" $\mu_{H_r}$ and $\mu_{H_i}$ which are subject to the following dynamics:
\beq \partial_t^2 \mu_{H_1} - c_{\mu_{H_1}}^2 \nabla^2 \mu_{H_1} + \lambda_{\mu_H} \frac{\partial \mu_{H_1}}{\partial t} = e_{H_1} \nonumber \eeq
\beq \partial_t^2 \mu_{H_{-1}} - c_{\mu_{H_{-1}}}^2 \nabla^2 \mu_{H_{-1}} + \lambda_{\mu_H} \frac{\partial \mu_{H_{-1}}}{\partial t} = e_{H_{-1}} \nonumber \eeq
\beq \partial_t^2 \mu_{H_i} - c_{\mu_{H_i}}^2 \nabla^2 \mu_{H_i} + \lambda_{\mu_H} \frac{\partial \mu_{H_i}}{\partial t} = e_{H_i} \eeq
\beq \partial_t^2 \mu_{H_{-i}} - c_{\mu_{H_{-i}}}^2 \nabla^2 \mu_{H_{-i}} + \lambda_{\mu_H} \frac{\partial \mu_{H_{-i}}}{\partial t} = e_{H_{-i}} \nonumber \eeq
Then, by the same reasoning as with $\psi$, we claim that if we introduce "receptor fields" they will be able to "listen" to everything emitted in the entire universe (since the speed of propagation of signal is very large, the time delays are negligible), which is the ultimate source of non-locality. Similarly to previous section, we define them based on driven harmonic oscillators, as follows:
\beq \partial_t^2 r_{H_1} (\vec{x}, t) = - \omega^2 (u_1 (\vec{x}), u_2 (\vec{x})) r_{H_1} (\vec{x}, t) - \lambda_{r_{H_1}} \partial_t r_{H_1} (\vec{x}, t) + \mu_{H_1} (\vec{x}, t) \nonumber \eeq
\beq \partial_t^2 r_{H_{-1}} (\vec{x}, t) = - \omega^2 (u_1 (\vec{x}), u_2 (\vec{x})) r_{H_{-1}} (\vec{x}, t) - \lambda_{r_{H_{-1}}} \partial_t r_{H_{-1}} (\vec{x}, t) + \mu_{H_{-1}} (\vec{x}, t) \eeq
\beq \partial_t^2 r_{H_i} (\vec{x}, t) = - \omega^2 (u_1 (\vec{x}), u_2 (\vec{x})) r_{H_i} (\vec{x}, t) - \lambda_{r_{H_i}} \partial_t r_{H_i} (\vec{x}, t) + \mu_{H_i} (\vec{x}, t) \nonumber \eeq
\beq \partial_t^2 r_{H_{-i}} (\vec{x}, t) = - \omega^2 (u_1 (\vec{x}), u_2 (\vec{x})) r_{H_{-i}} (\vec{x}, t) - \lambda_{r_{H_{-i}}} \partial_t r_{H_{-i}} (\vec{x}, t) + \mu_{H_{-i}} (\vec{x}, t) \nonumber \eeq
Finally, we would like to get rid of unwanted oscillations of $r_H$ by defining $H$ to be an \emph{amplitude} of the above oscillations. By the same arguments we used in the previous section, the way to do that without violating locality in time is through the following dynamics: 
\beq \partial_t^2 H_1 = a (e^{b(r_{H_1}^2 - H_1)} -1 ) \; ; \;  \partial_t^2 H_{-1} = a (e^{b(r_{H_{-1}}^2 - H_{-1})} -1 )\eeq
\beq \partial_t^2 H_i = a (e^{b(r_{H_i}^2 - H_i)} -1 ) \; ; \;  \partial_t^2 H_{-i} = a (e^{b(r_{H_{-i}}^2 - H_{-i})} -1 ) \nonumber \eeq
where $b$ is very large. 

\subsection*{7. Beables}

\subsection*{7.1 Review of D\"urr et el}

Up until this point we have shown how to model quantum field theory amplitudes in terms of functions \emph{only} over $\mathbb{R}^3$. In this chapter we will do the same with theory of measurement. We will follow similar philosophy in a sense that we will start out with global theory and then "convert" it into local one. In principle this can be done with any pilot wave model. However, in light of the fact that our configuration space is discrete, it is easiest to use the model of stochastic jumps due to D\"urr et el (see \cite{jumps}). Other models, on the other hand, would likely have to be artificially discretized before they can be "plugged into" our framework. For that reason, in this paper we will focus exclusively on D\"urr et el, while leaving the rest of the Pilot Wave models for future work. In this subsection, we will briefly outline the \emph{global} picture proposed by D\"urr et el, and then in the next subsection we will move on to "encoding" it into $\mathbb{R}^3$.

The key concept of Pilot Wave models is that a particle and a wave are completely separate substances. There is no such thing as "collapse of wave function". Instead, a wave evolves according to Schr\"odinger's equation (or some other rules of quantum mechanics) \emph{at all times}. At the same time, a particle is local \emph{at all times}, and it is being guided by a wave according to \emph{guidance equation}. In case of non-relativistic quantum mechanics, that equation is 
\beq \frac{d \vec{x}}{dt} = \frac{1}{m} \; \vec{\nabla} \; Im \; ln \; \psi, \eeq
but that equation might become increasingly more complicated for more general cases, specifically involving quantum fields. In all cases, however, the guidence equation has to be designed in such a way that if we don't have enough information about the particle, then the probability of finding it will end up being $\vert \psi \vert^2$.

Let us now translate the above into the language of multiparticle states. The notion of wave function generalizes to an assignment of probability amplitudes to all possible states. Schr\"odinger's equation generalizes to Feynman rules. And, finally, the notion of localized particle generalizes to one specific state. Thus, the claim of any Pilot Wave model is that, at any given time, we assign the probability amplitudes $\psi$ to \emph{all} states, and, at the same time, we have one specific state that evolves in time, guided by these amplitudes. 

One of the main challenges of this, however, is that the evolution of state involves creation and annihilation of particles, which is not a continuous process, while the above guidance equation is continuous. D\"urr et el decided to address this issue by replacing the velocity $\vec{v}$ with a probability of a jump, $\sigma (e, e')$ from the state $e$ to the state $e'$. This, however, violates determinism because, while we know the probability, the actual timing of each jump is random. Thus, in order to be as close as possible to determinism, they have claimed that \emph{between} the jumps our state evolves according to differentiable (and, therefore, deterministic) equation. That process is simply being "interrupted" by the discrete jumps.

Now, in order to calculate the probability of these jumps, we have to compute the time derivative of \emph{desired} probability of each state, based on quantum field theory. After that, a suitable Pilot Wave model is introduced in order to accommodate that desired probability. Now, suppose the evolving state is $\vert \Psi >$, and the probability of a state $\vert e >$ is $\psi (e) = <e \vert \Psi>$ Then the total current flowing into that state is given by 
\beq \frac{d}{dt} (\psi^* \psi) = 2 \; Re \; \Big( \psi^* \frac{d \psi}{dt} \Big) \eeq
Now, from evolution equation we know that 
\beq \frac{d \psi}{dt} = i <e \vert H \vert \Psi> = i \sum_{e'} <e \vert H \vert e'><e' \vert \Psi>. \eeq
Combining this with 
\beq \psi^* = < \Psi \vert e> \eeq
we get
\beq \frac{d}{dt} (\psi^* \psi) = 2 \; Im \; \sum_{e'} \; ( < \Psi \vert e> <e \vert H \vert e'><e' \vert \Psi> ) \eeq
This means that we would like a current from $\vert e' >$ to $\vert e>$ to be given by 
\beq j(e', e) = 2 \; Im ( < \Psi \vert e> <e \vert H \vert e'><e' \vert \Psi> ) \eeq
Now, that current is equal to the probability of a jump, multiplied by the probability density of $e'$ (namely, $\psi^* \psi = <e' \vert \Psi><\Psi \vert e'>$). Thus, in order to obtain the probability of the jump, we have to \emph{divide} the above expression by the latter. Thus, our first guess is
\beq \sigma (e', e) =^? 2 \; Im \; \frac{< \Psi \vert e> <e \vert H \vert e'><e' \vert \Psi>}{<e' \vert \Psi><\Psi \vert e'>} \eeq
Naively, this would give us a negative probability of jumps in direction opposite to the current. This, of course, wouldn't make sense. So, instead, we claim that the probability of the latter is $0$. We can do that aby introducing a function $x^{\dagger}$ which is equal to $x$ when $x>0$ and $0$ when $x<0$.   This gives us
\beq \sigma (e', e) = 2 \; Im \; \frac{(< \Psi \vert e> <e \vert H \vert e'><e' \vert \Psi>)^{\dagger}}{<e' \vert \Psi><\Psi \vert e'>} \eeq
As we previously said, however, according to D\"urr et el, the above is just \emph{part} of a dynamics. In particular, according to that model most of the time the system evolves according to a differentiable, deterministic equation. It is merely being \emph{interrupted} by stochastic jumps with the above probability. However, D\"urr et el also notices that continuum process, itself, might be simply a limit to the discrete. In particular, they quote Bell's suggestion of representing fermions in terms of the numbers of particles at every discrete lattice points. 

The ultimate reason why D\"urr et el tends to favor the option that does include continuous processes over the time intervals is, obviously, because it is more deterministic, although still not completely so. In our work, however, we will show how the jumps themselves can be modeled as deterministic processes. In the language of introduction, we are building a "computer" that "simulates" anything and everything we might want. We know that a computer is ran by differentiable, deterministic processes. Thus, as long as we will successfully simulate something, it will be continuous and deterministic, by default. In other words, the jumps are no longer "worse" than the continuous process.

 On the other hand, in light of the fact that our configuration space is discrete, modeling continuous process is very problematic; attempting to do so would probably require to discretize it first! For that reason, we will assume a Pilot Wave model that is being ran by above-described jumps \emph{alone}, without any continuous evolution. That, of course, is possible. After all, we don't have access to very small scales so whatever appears continuous to our eyes might, in reality, be discrete. 

\subsection*{7.2 Converting D\"urr et el into a local theory}

We are now ready to "convert" the non-local model of D\"urr into our "local" theory. In order to do that, we will introduce one single beable particle, whose position is $\vec{x} = \vec{b} (t) \in \mathbb{R}^3$. We will also introduce beable field $B \colon \mathbb{R}^3 \times \mathbb{R} \rightarrow \mathbb{R}$. Our goal is to find a dynamics that would guarantee us that $B(\vec{x}, t)$ is non-zero \emph{only} when $u_1 (\vec{x}) = u_1 (\vec{b} (t))$ and, in the latter case,  $B(\vec{x}, t)$ is roughly independent of the choice of $\vec{x}$.

If the above is done successfully, it guarantees us that the position of \emph{single} particle in $\mathbb{R}^3$ (which we will call b-particle) "encodes" a configuration of multiple particles that is realized at a given time. In particular, the b-particle is like a "lamp" that "illuminates" the particles that have approximately the same value of $u_1$, and doesn't illuminate anything else. Then, when our "lamp" moves, the values of $u_1$ the places it occupies change. Thus, the set of particles it "illuminates" changes as well, which leads to an illusion that they move around.

We would like to be able to find a local theory that imitates non-local behavior described in D\"urr et el (\cite{jumps}). In that paper, they claimed that within small interval $dt$, the probability of a "jump" between the state $\vert e>$ and $\vert e' >$ is $\sigma (e, e') dt$, where
\beq \sigma (e, e') = \frac{<\psi \vert e> <e \vert H \vert e'> <e' \vert \psi>}{\vert <\psi \vert e> \vert^2} \eeq
In order to mimic that result, we have to claim that, over some short period of time, the probability that b-particle would travel from $x$ to $x'$ is proportional to $\sigma (S_{u (x)}, S_{u (x')})$, where $S_{u_0}$ is a quantum state that can be "read off" from $u \approx u_0$. But then the question is: what is a \emph{local} mechanism for b-particle to "scan" the values of $u(\vec{x}')$ at various $\vec{x}'$ and "decide" where to move? 

Thankfully, from previous sections, we already have a clue of where to look at: we have two fields $u_1$ and $u_2$. We recall that, as far as quantum mechanical amplitudes were concerned, every point $\vec{x}$ can serve as a "transition" from the state $S_{u_1 (\vec{x})}$ to the state $S_{u_2 (\vec{x})}$. We will now introduce similar concept for beables. In other words, the point $\vec{x}'$ to where the b-particle transitions to, has to satisfy $u_1 (\vec{x}') \approx u_2 (\vec{x})$. The reason $u_1 (\vec{x}')$ appears to be random is that we \emph{only} know $u_1 (\vec{x})$ and \emph{not} $u_2 (\vec{x})$. After all, only the former can be "read off" from the configuration of the "visible" particles. 

Again, the above is non-local, so we need to come up with some local mechanism for that. Furthermore, we also have to come up with a way in which a particle "decides" to \emph{which} of the $\vec{x}' \in S_{u_2 (\vec{x})}$ to transition to, \emph{which} trajectory to take, etc. The mechanism that we propose is that a particle sends off signals that cause a "potential" to fall throughout the \emph{entire} $S_{u_2 (\vec{x})}$. After that, it is attracted to that same potential that it caused. Its specific trajectory is determined based on its initial location.

Before we proceed with this, we have to make sure that we enforce the probability $\sigma (e, e')$ defined above. We will first find a \emph{local} way of defining the \emph{desired} probability, $\sigma (\vec{x})$, and then afterwards propose a deterministic mechanism for that desired probability to be the actual one. In order to come up with desired probability, the information about $\psi (\vec{x}')$  (or, in other words, $\psi (S_{u_2 (x)})$) has to be available at $\vec{x}$. 

This can be done by furnishing point $\vec{x}$ with an "antenna" tuned to $u_2$ rather than $u_1$, and thus introducing another receptor fields, $r_{\psi'}$ and $\psi'$ that are being "read" from that antenna. Since the mechanism of obtaining $r_{\psi'} (\vec{x})$ and $\psi' (\vec{x})$ is the same as the one for $r_{\psi} (\vec{x}')$ and $\psi (\vec{x}')$, respectively, the former should approximate the latter. The good news is that the former represents fields at $\vec{x}$, and \emph{not} at $\vec{x}'$. This allows us to come up with a local expression for $\sigma (\vec{x}, \vec{x}')$.

Now, the mechanism of getting $r_{\psi'}$ and $\psi'$ is a carbon copy of the equations for $r_{\psi}$ and $\psi$, where $\psi$ is replaced with $\psi'$, $r_{\psi}$ is replaced with $r_{\psi'}$ and $u_1$ is replaced with $u_2$. However, we do \emph{not} replace $e_{\psi}$ with $e_{\psi'}$; and we do \emph{not} replace $\mu_{\psi}$ with $\mu_{\psi'}$ either. The reason for this is that we would like $r_{\psi'} (\vec{x})$ and $r_{\psi} (\vec{x}')$ to be "listening" to \emph{the same} thing, namely $\mu_{\psi}$ (and \emph{not} $\mu_{\psi'}$) in order to be equal.

As before, we need to have four sets of equations corresponding to indexes $1$, $-1$, $i$ and $-i$. Thus, the set of equations of $r_{\psi'}$ is 
\beq \partial_t^2 r_{\psi'_1} (\vec{x}, t) = - u_2 (\vec{x}) r_{\psi'_1} (t, \vec{x}) - \lambda_r \partial_0 r_{\psi'_1} (t, x) + \mu_{\psi_1} (t, x) \nonumber \eeq
\beq \partial_t^2 r_{\psi'_{-1}} (\vec{x}, t) = - u_2 (\vec{x}) r_{\psi'_{-1}} (t, \vec{x}) - \lambda_r \partial_0 r_{\psi'_{-1}} (t, x) + \mu_{\psi_{-1}} (t, x) \nonumber \eeq
\beq \partial_t^2 r_{\psi'_i} (\vec{x}, t) = - u_2 (\vec{x}) r_{\psi'_i} (t, \vec{x}) - \lambda_r \partial_0 r_{\psi'_i} (t, x) + \mu_{\psi_i} (t, x) \eeq
\beq \partial_t^2 r_{\psi'_{-i}} (\vec{x}, t) = - u_2 (\vec{x}) r_{\psi'_{-i}} (t, \vec{x}) - \lambda_r \partial_0 r_{\psi'_{-i}} (t, x) + \mu_{\psi_{-i}} (t, x) \nonumber \eeq
Our way of obtaining $\psi'$ from $r_{\psi'}$ is the same as the way we obtained $r_{\psi}$ from $\psi$, which is
\beq \partial_0^2 \psi'_1 = a (e^{b(r_{\psi'_1}^2 - \psi_1)} -1) \; ; \; \partial_0^2 \psi'_{-1} = a (e^{b(r_{\psi'_{-1}}^2 - \psi'_{-1})} -1) \nonumber \eeq
\beq \partial_0^2 \psi'_i = a (e^{b(r_{\psi'_i}^2 - \psi'_i)} -1) \; ; \; \partial_0^2 \psi'_{-i} = a (e^{b(r_{\psi'_{-i}}^2 - \psi'_{-i})} -1)  \eeq
Now, as we have just mentioned, we interpret the values of $\psi$ and $\psi'$ as 
\beq \psi (x) = < \psi \vert S_{u_1 (\vec{x})} > \; ; \; \psi' (x) = < \psi \vert S_{u_2 (\vec{x})} > \eeq
Furthermore, as we explained in previous sections, the Hamiltonian is encoded in $H(\vec{x})$ via
\beq <S_{u_1} (\vec{x}) \vert H \vert S_{u_2} (\vec{x})> = H (\vec{x}, t) \eeq
Substituting these into the equation proposed by D\"urr et el, we obtain
\beq \sigma (t; S_{u_1 (\vec{x})}, S_{u_2 (\vec{x})}) = \sigma (\vec{x}, t) \eeq
where
\beq \sigma (\vec{x}, t) = \frac{Im (\psi (\vec{x}, t) \psi' (\vec{x}, t) H(\vec{x}, t))^{\dagger}}{\vert \psi (x) \vert^2}. \eeq
The above gives us the \emph{desired} probability of the transition of the b-particle. Now we would like to discuss the mechanism of that transition as well as how that desired probability is enforced. As we said before, we would like our b-particle to emit a signal that will cause the potential to fall throughout $S_{u_2 (\vec{x})}$. Thus, we would like to find a \emph{deterministic} mechanism of emission of a signal in such a way that it \emph{appears} to occur with probability $\sigma (\vec{x}, t)$. We do that by introducing \emph{a-particles} whose density is proportional to $\sigma (\vec{x}, t)$ (while there is only one b-particle there are many a-particles), and the signal is emitted if and only if our b-particle collides with one of the a-particles. 

Since the particles have zero size, they have zero probability of collision. Thus, we have to define a field $\rho$ that surrounds each particle, and by "collision" we really mean the overlap of these fields (intuitively, this is equivalent to saying that particles have size, and $\rho$ is the density of matter distribution that makes up these particles). We will denote these fields, corresponding to particles a and b, as $\rho_a$ and $\rho_b$, respectively. It doesn't matter what specific expression we choose for them as long as it meets the above property. For definiteness, we will postulate 
\beq \rho_a (\vec{x}, t) = \sum_{k=1}^{n_a} \int_{t'<t} \delta (c_{\rho_a} (t-t') - d (\vec{x}, \vec{a}_k (t'))) \nonumber \eeq
\beq \rho_b (\vec{x}, t) = \int_{t'<t} \delta (c_{\rho_b} (t-t') - d (\vec{x}, \vec{b} (t'))) \eeq
Here, we put a summation sign for a-particles but \emph{not} for b, because we assume that there is only one b particle, which corresponds to a single state defined by a beable. Now, in order to enforce the desired density of $a$-particles, we come up with a dynamics that assures us that their velocity is inversely proportional to the desired density. Thus, the larger the desired density is, the slower they move, and thus they spend more time in that region.  In order for the direction of their velocity to be specified as well, we will use \emph{acceleration} to attain the desired velocity. The direction of that acceleration is parallel to direction of velocity at any given time:
\beq \frac{d^2 \vec{a}_k}{dt^2} = k \frac{d \vec{a}_k}{dt} \Big(\sigma (\vec{a}_k, t) - \Big\vert \frac{d \vec{a}_k}{dt} \Big\vert \Big) \eeq
In other words, from the space-alone perspective, a-particles always move along geodesics, with varying speeds. But in light of the compactness of the universe, these particles come back moving in a very different direction after having circled the latter. This provides mechanism of randomness of their velocity directions. 

We will now have to introduce a mechanism of generating the desired potential when the overlap between $\rho_a$ and $\rho_b$ occurs. As always, we will do that through emission field $e_V$, messenger field $\mu_V$, reception field $r_V$ and finally a potential $V$. This is a bit tricky though. Suppose b-particle, after collision of a-particle, was "sent" to some region $S_{u_2 (\vec{x})}$. But on its way there it was crossing a region $S_{u_3}$, and there it underwent collision with another a-particle that sent it to $S_{u_4}$. In order to avoid this situation, we have to make sure that the "source" of $e_V$ is large only if the velocity of a particle is small (thus, the particle can only be "sent" to a new destination after it has "arrived" to the previous one, and, therefore, slowed down). Since the interaction occurs through overlap of $\rho$-fields rather than actual collision of particles, the "velocity" relevent to this interaction is the "current" of $\rho_b$, defined as 
\beq \vec{j}_b (\vec{x}, t) = \int_{t'<t} \frac{d \vec{b}}{dt} \delta (c_{\rho_b} (t-t') - d (\vec{x}, \vec{b} (t'))). \eeq
The reason we have only introduced $\vec{j}_b$ and not $\vec{j}_a$ is that our goal is to avoid the interaction when b-particle is in transition. The latter is characterized by $\vec{j}_b$ being large. The values of $\vec{j}_a$ are independent of that and, therefore, should be irrelevant. In the above expression we used $c_{\rho}$ instead of coming up with a separate constant $c_j$, and that is for a good reason. If the values of $c_{\rho}$ and $c_j$ were different, then the two fields would have different ranges of interaction. Now, if the range of $\vec{j}_b$ was smaller than the range of $\rho_b$, then some distance away from b-particle it would appear that the latter is stationary. Thus, in that region a-b interaction will occur "by mistake". Since we don't want that, we make sure that $c_{j_b} = c_{\rho_b}$.

Apart from wanting emission to be small whenever $\vec{j}_b$ is large, we also would like it to be small when the magnitude of a "potential", $\vert V \vert$, is large, as well. Suppose a particle makes a transition from $\vec{x}_1 \in S_{u_1}$ to $\vec{x}_2 \in S_{u_2}$ and then, later, it makes another transition from $\vec{x}_3 \in S_{u_2}$ to $\vec{x}_4 \in S_{u_3}$ (the reason we have four $\vec{x}$-s is that a particle can move while in $S_{u_2}$, as will be discussed later). Now, the fact that the transition from $\vec{x}_1$ to $\vec{x}_2$ occurred, indicates that there was a fall in potential throughout $S_{u_2}$. Now, suppose that potential "didn't have time" to return to $0$ after the second transition. Then, the fact that we \emph{still} have potential well throughout $S_{u_2}$ might lead to a particle transitioning to $\vec{x}_5 \in S_{u_2}$. We do not want that. For that reason, we would like b particle to "wait" until the potential becomes approximately $0$ before making another transition. Thus, we would like $\vert V \vert$ to have damping effect on emission, on top of the dumping effect of $\vert{j}_b \vert$. It is important to indicate, though, that even though $\vert V \vert$ is \emph{approximately} $0$ throughout $S_{u_2}$, it is not exactly $0$ yet; thus, particle \emph{still} spends most of the time in that region.

We are now ready to describe the emission field $e_V$ produced by a-b interaction. As before, since emission field is non-propagating, its equation does not involve any space derivatives. But, at the same time, in order to avoid $\delta$-function singularities it is defined as a function both of space and time. Its equation of motion is the one of dumped harmonic oscillator, whose source is proportional to $\rho_a \rho_b$. In order for the source to be small whenever either $\vec{j}_b$ or $\vert V \vert$ is large, we divide it by $1+ N_b j_b + N_V \vert V \vert$, where $N_b$ and $N_V$ are some large constants. This gives us 
\beq \partial_t^2 e_V (t, \vec{x}) = -u_2 (t, \vec{x}) e_V (t, \vec{x}) - \lambda_{e_V} \partial_t e_V (t, \vec{x}) + \frac{\rho_a (t, \vec{x}) \rho_b (t, \vec{x})}{1+N_b j_b (t, \vec{x})+ N_V \vert V (\vec{x}, t) \vert} \eeq
In order to obtain the desired fall in potential throughout $S_{u_2 (\vec{x})}$, we would like to introduce a messenger signal, $\mu_V$ that is being emitted by $e_V$. Its equation of motion is given by 
\beq \partial_t^2 \mu_V (t, \vec{x}) - c_{\mu_V}^2 \nabla^2 \mu_V (t, \vec{x}) + \lambda_{\mu_V} \mu_V  (t, \vec{x}) = \mu_V (t, \vec{x}) \eeq
Finally, the desired value of $V$ is produced through the "reception" of that signal. For the same reasons as explained in previous sections, the reception is a two step process that involves the oscillating reception field $r_V$ and then its conversion into non-oscillating $V$: 
\beq \partial_t^2 r_V (t, \vec{x}) = -u_1 (t, \vec{x}) r_V(t, \vec{x}) - \lambda_{r_V} \partial_t r_V (t, \vec{x}) + \mu_V (t, \vec{x}) \eeq
\beq \partial_t^2 V(t, \vec{x}) = k_V (e^{l_V (r_V^2 - V)} -1 ) \eeq
In the above expressions, $\lambda_{r_V}$ and $k_V$ have to be sufficiently large, so that after a transition of the b particle is completed, the potential will disappear soon enough; after all, we don't want our particle to "wait" for the next transition for too long, since that would distort the stochastic nature of the desired jump. At the same time, however, these coefficients should not be too large either. After all, if the potential "dies out" too fast, the b-particle might not complete a transition. Since its path is continuous, it will be "stuck" at some point $\vec{x}''$ for which $u(\vec{x}'') = u_3$. Given that the $u$-field fluctuates quite fast, $u_3$ will have nothing to do with either $u_1$ or $u_2$, which would lead to a transition to unwanted quantum state.

Thankfully, it is quite easy to make sure that the potential dies out neither too slowly nor too fast. As we have explained in previous sections, the $u$ field has to be designed in such a way that small enough neighborhood of every point intersects every single $u_k$. If such is the case, it will take very short time for a particle to make a transition. Therefore, even if potential dies out very fast, it might still \emph{not} be \emph{too} fast. Thus, the answer to our dilemma is for the above mentioned constants to be very large but, at the same time, be smaller than some power of the inverse of the scale of fluctuations of $u$ (the latter is much larger than any large number we would normally think of). 

There is one more thing to take care of. Since the regions $S_{u_1}$ and $S_{u_2}$ do not "touch" each other, the b particle will not be able to find out a direction in which it has to travel based on $\vec{\nabla} V$. We will enforce the desired transition by making sure that the velocity of the b particle is large as long as $\vert V \vert$ is small. Thus, a particle will keep moving, until it "accidentally" reaches a region of large $\vert V \vert$, where it will "stop". Thus, we would like the velocity of the b particle to be $1/(1+M_V V)$, where $M_V$ is some very large number. In order for the direction of the velocity to be well defined, we will use our earlier trick and impose a condition on \emph{acceleration}; namely, that a particle accelerates in a direction \emph{parallel} to its velocity until its speed reaches a desired value:
\beq \frac{d^2 \vec{b}}{dt^2} = k_b \frac{d \vec{b}}{dt} \Big(\frac{1}{1+ N_V V} - \Big\vert \frac{d \vec{b}}{dt} \Big\vert \Big)  \eeq
The above completes the description of dynamics of b-particle. Thus, we have found a way in which b-particle "jumps" into the desired $S_{u_k}$. But it only occupies one point in that set. We would like to propose a mechanism by which from that one point it "illuminates" the \emph{entire} $S_{u_k}$, thus making all of the "electrons" (and other stationary particles) that happen to reside in $S_{u_k}$ visible. In previous versions of this paper, we have done this by introducing a new field, $B$, which is produced by a $b$ particle. However, right now I realized that the potential $V$ has essentially the same properties; thus, based on Ocam's razor, it is better to simply use $V$. But, in order for the notation to be consistent with previous versions of the paper, we will still use $B$ and simply define it as $B(\vec{x}) = V (\vec{x})$, This does not agree with more complicated definition of $B$ in previous versions, but it serves the same purpose. 

Lets finally explicitly write the way $B$ field "illuminates" our states. As we explained in the beginning, the particles we are ultimately interested in (electrons, protons, etc) are all stationary. They fall into some regions $S_{u_k}$, and they are either visible or invisible, based on the value of $\psi$ in that region. In light of the fact that we have introduced Pilot Wave model, we have to replace $\psi$ with $B$ as the ultimate definition of reality. Thus, our particles are either visible or invisible based on value of $B$. Now, as we mentioned before, each particle has two short-acting $\rho$ fields: for a particle of type $k$ they are $\rho_{k1}$ and $\rho_{k2}$. They represent  a "density" of the version of that particle with a size. These fields are independent of how much (or how little) a particle is being "illuminated" by $B$. Now, in order to take that illumination into account, we will postulate that the "observed" density of the particles is $\tilde{\rho}_k$, and is defined to be 
\beq \tilde{\rho}_k (\vec{x}, t) = \rho_{k1} (\vec{x}) B (\vec{x}, t) \eeq
On the right hand side of the above expression I have chosen $\rho_{k1}$ instead of $\rho_{k2}$ because for a particle located at $\vec{x} \in S_i \cap T_j$, $\rho_1$ is felt withing the $S_i$-neighborhood of $\vec{x}$ while $\rho_2$ is felt within $T_j$-neighborhood of $x$. Thus, in light of the fact that the intensity of $B$ field is a function of $S$-s as opposed to $T$-s, $\rho_1$ is more appropriate choice. Intuitively, $\rho_{k1}$ is a sum of the finite-width versions of $\delta$-functions, and the $B$ is a coefficient in front of the sum that "picks out" which of the terms we "see" and which we do not. In principle, a reader that prefers the situation where particles have $0$ size is free to replace $\rho_{k1}$ in the above expression with iteral $\delta$-functions. 

\subsection*{8. Conclusion}

In this paper we have presented a way to "encode" quantum field theory amplitudes into three dimensional space. Then, later, we used similar techniques to "convert" Pilot Wave model due to D\"urr et el from configuration space to ordinary $\mathbb{R}^3$ as well. Apart from getting rid of non-local concept of configuration space, we found a way to make sure that the "jumps" postulated by these authors are continuous, and, therefore, deterministic.

One obvious short coming of this work is that the approach is very much "forced". We basically know the kind of answers we want, and we design a "machine" that simulates these answers, while reminding ourselves all along that a classically-working "machine" is local by default. In some sense this is cheating since any physical concept we "don't like" can be explained away by the "machine" that simulates it. Besides, this work implies a fractal-type structure: the quantum arises out of classical while it is still true that on larger scales the classical arises out of quantum. This might lead to a question: why the coincidence?

On the other hand, however, one might argue that whenever we conceptualize anything we are doing that based on the concept we are used to. In fact, when we think of space, we imagine that we have eyes to see it; or when we think of time, we imagine some changes in our brain that make us feel time. Thus, the question "why did fractals arise" can be asked in that level, too. It is also important to note that \emph{if} a cave man was to see a computer, without knowing what it is, he, too, might think of some "quantum" concepts similar to configuration space; and, if someone were to tell him about the complicated circuits involved, he would also say that this explanation is "forced". But that doesn't make him right.

There are also some unresolved issues on a more technical level. In particular, this work was using an "imaginary" version of quantum field theory, where particles have size, and the "vortexes" are replaced by the functions of the "matter density". No attempt to formulate such theory was made. Instead, it was simply \emph{assumed} that the configuration space version of that theory exists, and we were focused on converting that \emph{imaginary} theory into the language of ordinary space. For the future research, it is important to fill in that gap.

\subsection*{A. Appendix: Driven harmonic oscillator in classical dynamics}

Throughout this paper, the concept of resonances was extensively used that is simply borrowed from the dynamics of classical harmonic oscillator. Therefore, for the convenience of the reader, in this section we will review the standard treatment of driven harmonic oscillaor. The same can also be read in a number of classical dynamics textbooks, including \cite{phys105}. 

Consider a harmonic oscillator with internal frequency $\omega_0$, driven by sinusoidal external force of frequency $\omega$. The equation of motion of this oscillator is given by 
\beq m \frac{d^2 x}{dt^2} = sin \; (\omega (t-t_0)) - m \omega_0^2 x - \lambda \frac{dx}{dt}. \eeq
In order to solve the above equation it is easier to first solve 
\beq m \frac{d^2 x}{dt^2} = e^{i \omega (t-t_0)} - m \omega_0^2 x - \lambda \frac{dx}{dt} \eeq
and then take an appropriate linear combination of its solutions. This equation can be re-expressed as
\beq m \frac{d^2 x}{dt^2} = A e^{i \omega t} - m \omega_0^2 x - \lambda \frac{dx}{dt}\eeq
where $A = e^{-i \omega t_0}$. We will try a solution of the form 
\beq x = B e^{i \omega t} \eeq
By substituting it into above equation, it becomes
\beq - m \omega^2 B = A - m \omega_0^2 B - i \omega \lambda B \eeq
which gives us
\beq B = \frac{A}{m(\omega_0^2 - \omega^2)+ i \omega \lambda } \eeq
and, therefore,
\beq x = \frac{A e^{i \omega t}}{m(\omega_0^2 - \omega^2) + i \omega \lambda } \eeq
By substituting $A = e^{-i \omega t_0}$, we obtain
\beq x = \frac{e^{i \omega (t-t_0)}}{m(\omega_0^2 - \omega^2)+i \omega \lambda } \eeq
We now return to the original equation,
\beq m \frac{d^2 x}{dt^2} = sin \; (\omega (t-t_0)) -m \omega_0^2 x - \lambda \frac{dx}{dt} \eeq
The solution to this equation is given by linear combination,
\beq x = \frac{x_1 - x_2}{2i} \eeq
where $x_1$ and $x_2$ satisfy the following equations:
\beq m \frac{d^2 x_1}{dt^2} = e^{i \omega (t-t_0)} - m \omega_0^2 x_1 - \lambda \frac{dx_1}{dt} \eeq
and
\beq m \frac{d^2 x_2}{dt^2} = e^{- i \omega (t-t_0)} - m \omega_0^2 x_2 - \lambda \frac{dx_1}{dt} \eeq
Thus, $x_1$ can be read off from the solution we have just found, while $x_2$ can be obtained by replacing $\omega$ with $- \omega$. Their linear combination gives
\beq x = \frac{1}{2i} \Big( \frac{e^{i \omega (t-t_0)}}{m(\omega_0^2 - \omega^2) + i \omega \lambda } - \frac{e^{-i \omega (t-t_0)}}{ m (\omega_0^2 - \omega^2) + i \omega \lambda } \Big) \eeq
which, after some algebra becomes
\beq x = \frac{m (\omega_0^2 - \omega^2) \;  sin \; ( \omega (t-t_0)) - \omega \lambda \; cos \; (\omega (t-t_0))}{m (\omega_0^2 - \omega^2)^2 + \omega^2 \lambda^2} \eeq
By using the identities 
\beq  cos \;  \Big( tan^{-1} \frac{\omega \lambda}{m^2 (\omega_0^2  - \omega^2)} \Big) = \frac{m(\omega_0^2 - \omega^2)}{\sqrt{m^2 (\omega_0^2 - \omega^2)^2 + \omega^2 \lambda^2}} \nonumber \eeq
\beq  sin \;  \Big( tan^{-1} \frac{\omega \lambda}{m^2 ( \omega_0^2 - \omega^2)} \Big) = \frac{\omega \lambda}{\sqrt{m^2 (\omega_0^2 - \omega^2)^2 + \omega^2 \lambda^2}}  \eeq
the expression for x becomes
\beq x = \frac{ cos \;  \Big( tan^{-1} \frac{\omega \lambda}{m (\omega_0^2 - \omega^2)} \Big) \; sin \; (\omega (t-t_0))  - sin  \;  \Big( tan^{-1} \frac{\omega \lambda}{m (\omega_0^2 - \omega^2)} \Big) \; cos \; (\omega (t-t_0))}{\sqrt{m^2 (\omega_0^2- \omega^2)^2 + \omega^2 \lambda^2}}. \eeq
By using the expression of the sine of the difference, this evaluates to
\beq x = \frac{sin \; (\omega (t- t_0 - t_1))}{\sqrt{m^2 (\omega_0^2 - \omega^2)^2 + \omega^2 \lambda^2}} \eeq
where
\beq  t_1 = \frac{1}{\omega} tan^{-1} \frac{\omega \lambda}{m^2 (\omega_0^2 - \omega^2)} \eeq
The above solution, while correct, is not complete. After all, we can easily choose initial conditions on position and velocity that do not meet the above equation. This problem is fixed by adding a \emph{complimentary solution} corresponding to a free oscillator,   
\beq m \frac{d^2 x_c}{dt^2} = -m \omega_0^2 x_c - \lambda \frac{dx_c}{dt}. \eeq
From linearity, it follows that if $x$ satisfies the equation of driven harmonic oscillator, so does $x+x_c$. The solution that we have obtained earlier is called \emph{particular solution}, and denoted by $x_p$. Thus, a general solution is of the form 
\beq x= x_c + x_p \eeq
To obtain $x_c$, we again look for the solution of the form
\beq x_c = b e^{at} \eeq
where $a$ can be any complex number (which is why we put $e^{at}$ instead of $e^{iat}$). Substituting this into the above differential equation, we get
\beq m a^2 + \lambda a  + m \omega_0^2 = 0 \eeq
which gives us
\beq a = \frac{-\lambda \pm \sqrt{\lambda^2 -4m^2 \omega_0^2}}{2m}. \eeq 
If we assume $ \lambda < 4k$, this becomes 
\beq a = \frac{-\lambda \pm i \sqrt{4m^2 \omega_0^2- \lambda^2}}{2m}. \eeq
Since we would like $x_c$ to be real, we express it as a linear combination of the two solutions:
\beq x_c = \frac{x_{c0}}{2} \Big(e^{\frac{-\lambda + \sqrt{\lambda^2 -4m^2 \omega_0^2}}{2}(t-t_0)} + e^{\frac{-\lambda - \sqrt{\lambda^2 -4m^2 \omega_0^2}}{2}(t-t_0)} \Big), \eeq
which easily evaluates to
\beq x_c = x_{c0} e^{-\frac{\lambda}{2} (t-t_{c0})} \; cos \; \Big( \frac{\sqrt{4m^2 \omega_0^2-\lambda^2}}{2} (t-t_0) \Big). \eeq
Thus, the complete solution is 
\beq x = \frac{sin \; (\omega (t- t_0 - t_1))}{\sqrt{m^2 (\omega_0^2 - \omega^2)^2 + \omega^2 \lambda^2}} +  x_{c0} e^{-\frac{\lambda}{2} (t-t_{c0})} \; cos \; \Big( \frac{\sqrt{4m^2 \omega_0^2-\lambda^2}}{2} (t-t_0) \Big), \eeq
where
\beq  t_1 = \frac{1}{\omega} tan^{-1} \frac{\omega \lambda}{m^2 (\omega_0^2 - \omega^2)} \eeq
However, due to $e^{-\frac{\lambda}{2}} (t - t_{c0})$ factor, the complimentary solution dies out in time. Thus, after enough time passes, only particular solution, 
\beq x_p =  \frac{sin \; (\omega (t- t_0 - t_1))}{\sqrt{m^2 (\omega_0^2 - \omega^2)^2 + \omega^2 \lambda^2}} \eeq
survives. The violation of time reversal symmetry is due to the fact that the velocity with a minus sign (as opposed to plus sign) enters into the acceleration. Notably, if $\lambda$ is very small, the amplitude corresponding to $\omega = \omega_0$ is very large. This phenomenon is called \emph{resonance}, and $\omega_0$ is called \emph{resonance frequency}.

%\newpage

\end{document}